%% file: main.tex
\pdfoutput=1

\documentclass[12pt,a4paper]{article}

\usepackage{ifthen} 
\newboolean{pdflatex}
\setboolean{pdflatex}{true} 

\newboolean{articletitles}
\setboolean{articletitles}{true} 

\newboolean{uprightparticles}
\setboolean{uprightparticles}{false} 

\def\paperauthors{LHCb collaboration} 
\def\paperasciititle{Search for the doubly charmed baryon Xicc+} 
\def\papertitle{Search for the doubly charmed baryon \Xiccp} 
\def\paperkeywords{{High Energy Physics}, {LHCb}} 
\def\papercopyright{\the\year\ CERN for the benefit of the LHCb collaboration} 
\def\paperlicence{CC-BY-4.0 licence}
\def\paperlicenceurl{https://creativecommons.org/licenses/by/4.0/}

\input{preamble}
\usepackage{longtable} 

\begin{document}

\renewcommand{\thefootnote}{\fnsymbol{footnote}}
\setcounter{footnote}{1}

\input{title-LHCb-PAPER}


\renewcommand{\thefootnote}{\arabic{footnote}}
\setcounter{footnote}{0}

\cleardoublepage

\pagestyle{plain} 
\setcounter{page}{1}
\pagenumbering{arabic}


%

\input{introduction}

\input{detector}

\input{selection}

\input{yield}

\input{efficiency}

\input{systematics}

\input{upper_limit}

\input{conclusions}

\input{acknowledgements}

\clearpage
\addcontentsline{toc}{section}{References}
\bibliographystyle{LHCb}
\bibliography{main}

\newpage
\input{LHCb_Authorship_30-Jul-2019}

\end{document}

%% file: preamble.tex
\usepackage[top=1in, bottom=1.25in, left=1in, right=1in]{geometry}

\usepackage{microtype}
\usepackage{lineno}  
\usepackage{xspace} 
\usepackage{caption} 

\usepackage{graphicx}  
\usepackage{color}
\usepackage{colortbl}
\graphicspath{{./figs/}} 
\DeclareGraphicsExtensions{.pdf,.PDF,png,.PNG}

\usepackage{amsmath} 
\usepackage{amssymb}
\usepackage{amsfonts}
\usepackage{upgreek} 

\newcommand*\patchAmsMathEnvironmentForLineno[1]{%
\expandafter\let\csname old#1\expandafter\endcsname\csname #1\endcsname
\expandafter\let\csname oldend#1\expandafter\endcsname\csname
end#1\endcsname
 \renewenvironment{#1}%
   {\linenomath\csname old#1\endcsname}%
   {\csname oldend#1\endcsname\endlinenomath}%
}
\newcommand*\patchBothAmsMathEnvironmentsForLineno[1]{%
  \patchAmsMathEnvironmentForLineno{#1}%
  \patchAmsMathEnvironmentForLineno{#1*}%
}
\AtBeginDocument{%
\patchBothAmsMathEnvironmentsForLineno{equation}%
\patchBothAmsMathEnvironmentsForLineno{align}%
\patchBothAmsMathEnvironmentsForLineno{flalign}%
\patchBothAmsMathEnvironmentsForLineno{alignat}%
\patchBothAmsMathEnvironmentsForLineno{gather}%
\patchBothAmsMathEnvironmentsForLineno{multline}%
\patchBothAmsMathEnvironmentsForLineno{eqnarray}%
}

\usepackage{hyperxmp}

\usepackage[pdftex,
            pdfauthor={\paperauthors},
            pdftitle={\paperasciititle},
            pdfkeywords={\paperkeywords},
            pdfcopyright={Copyright (C) \papercopyright},
            pdflicenseurl={\paperlicenceurl}]{hyperref}

\usepackage[colorinlistoftodos,textsize=scriptsize]{todonotes}

\usepackage[all]{hypcap} 

\input{lhcb-symbols-def} 

\input{fixed-symbols-def} 
\input{user-symbols-def}

\usepackage{siunitx}
\usepackage{cite} 
\usepackage{mciteplus}

%% file: lhcb-symbols-def.tex
\usepackage{xspace} 
\usepackage{upgreek}

\def\lhcb   {\mbox{LHCb}\xspace}

\def\babar  {\mbox{BaBar}\xspace}
\def\belle  {\mbox{Belle}\xspace}

\def\MagUp {\mbox{\em Mag\kern -0.05em Up}\xspace}

\ifthenelse{\boolean{uprightparticles}}%
{

 \def\Ppi         {\ensuremath{\uppi}\xspace}

 \def\PDelta      {\ensuremath{\Delta}\xspace}                 
 \def\PXi         {\ensuremath{\Xi}\xspace}                 
 \def\PLambda     {\ensuremath{\Lambda}\xspace}                 
 \def\PSigma      {\ensuremath{\Sigma}\xspace}                 
 \def\POmega      {\ensuremath{\Omega}\xspace}                 
 \def\PUpsilon    {\ensuremath{\Upsilon}\xspace}

 \def\PB      {\ensuremath{\mathrm{B}}\xspace}                 
                  
 \def\PD      {\ensuremath{\mathrm{D}}\xspace}

 \def\PK      {\ensuremath{\mathrm{K}}\xspace}

 \def\PW      {\ensuremath{\mathrm{W}}\xspace}

 \def\Pb      {\ensuremath{\mathrm{b}}\xspace}                 
 \def\Pc      {\ensuremath{\mathrm{c}}\xspace}                 
 \def\Pd      {\ensuremath{\mathrm{d}}\xspace}

 \def\Pi      {\ensuremath{\mathrm{i}}\xspace}

 \def\Pp      {\ensuremath{\mathrm{p}}\xspace}

 \def\Ps      {\ensuremath{\mathrm{s}}\xspace}                 
                  
 \def\Pu      {\ensuremath{\mathrm{u}}\xspace}

 \def\thebaroffset{0.0em}
}
{

 \def\Ppi         {\ensuremath{\pi}\xspace}

 \mathchardef\PDelta="7101
 \mathchardef\PXi="7104
 \mathchardef\PLambda="7103
 \mathchardef\PSigma="7106
 \mathchardef\POmega="710A
 \mathchardef\PUpsilon="7107
                  
 \def\PB      {\ensuremath{B}\xspace}                 
                  
 \def\PD      {\ensuremath{D}\xspace}

 \def\PK      {\ensuremath{K}\xspace}

 \def\PW      {\ensuremath{W}\xspace}

 \def\Pb      {\ensuremath{b}\xspace}                 
 \def\Pc      {\ensuremath{c}\xspace}                 
 \def\Pd      {\ensuremath{d}\xspace}

 \def\Pi      {\ensuremath{i}\xspace}

 \def\Pp      {\ensuremath{p}\xspace}

 \def\Ps      {\ensuremath{s}\xspace}                 
                  
 \def\Pu      {\ensuremath{u}\xspace}

 \def\thebaroffset{0.18em}
}
\newcommand{\offsetoverline}[2][\thebaroffset]{\kern #1\overline{\kern -#1 #2}}%

\makeatletter
\ifcase \@ptsize \relax
  \newcommand{\miniscule}{\@setfontsize\miniscule{4}{5}}
\or
  \newcommand{\miniscule}{\@setfontsize\miniscule{5}{6}}
\or
  \newcommand{\miniscule}{\@setfontsize\miniscule{5}{6}}
\fi
\makeatother

\DeclareRobustCommand{\optbar}[1]{\shortstack{{\miniscule (\rule[.5ex]{1.25em}{.18mm})}
  \\ [-.7ex] $#1$}}

\def\W      {{\ensuremath{\PW}}\xspace}

\def\uquark    {{\ensuremath{\Pu}}\xspace}

\def\dquark    {{\ensuremath{\Pd}}\xspace}

\def\squark    {{\ensuremath{\Ps}}\xspace}

\def\cquark    {{\ensuremath{\Pc}}\xspace}

\def\bquark    {{\ensuremath{\Pb}}\xspace}

\def\pion   {{\ensuremath{\Ppi}}\xspace}

\def\pip    {{\ensuremath{\pion^+}}\xspace}
\def\pim    {{\ensuremath{\pion^-}}\xspace}

\def\kaon    {{\ensuremath{\PK}}\xspace}

\def\KorKbar {\kern \thebaroffset\optbar{\kern -\thebaroffset \PK}{}\xspace}

\def\Km      {{\ensuremath{\kaon^-}}\xspace}

\def\D       {{\ensuremath{\PD}}\xspace}

\def\DorDbar {\kern \thebaroffset\optbar{\kern -\thebaroffset \PD}\xspace}

\def\Dp      {{\ensuremath{\D^+}}\xspace}

\def\B       {{\ensuremath{\PB}}\xspace}

\def\BorBbar {\kern \thebaroffset\optbar{\kern -\thebaroffset \PB}\xspace}

\def\Bd      {{\ensuremath{\B^0}}\xspace}

\def\BdorBdbar {\kern \thebaroffset\optbar{\kern -\thebaroffset \Bd}\xspace}

\def\Bs      {{\ensuremath{\B^0_\squark}}\xspace}

\def\BsorBsbar {\kern \thebaroffset\optbar{\kern -\thebaroffset \Bs}\xspace}

\def\Y#1S{\ensuremath{\PUpsilon{(#1S)}}\xspace}

\def\proton      {{\ensuremath{\Pp}}\xspace}

\def\Lz          {{\ensuremath{\PLambda}}\xspace}

\def\LorLbar     {\kern \thebaroffset\optbar{\kern -\thebaroffset \PLambda}\xspace}

\def\Xires       {{\ensuremath{\PXi}}\xspace}

\def\Omegares    {{\ensuremath{\POmega}}\xspace}

\def\Lc          {{\ensuremath{\Lz^+_\cquark}}\xspace}

\def\Xicp        {{\ensuremath{\Xires^+_\cquark}}\xspace}

\def\Xicc        {{\ensuremath{\Xires_{\cquark\cquark}}}\xspace}

\def\Xiccp       {{\ensuremath{\Xires^+_{\cquark\cquark}}}\xspace}
\def\Xiccpp      {{\ensuremath{\Xires^{++}_{\cquark\cquark}}}\xspace}

\def\Omegacc     {{\ensuremath{\Omegares^+_{\cquark\cquark}}}\xspace}

\def\Lb           {{\ensuremath{\Lz^0_\bquark}}\xspace}

\def\BF         {{\ensuremath{\mathcal{B}}}\xspace}

\def\to                 {\ensuremath{\rightarrow}\xspace}

\def\eps   {{\ensuremath{\varepsilon}}\xspace}

\def\AT#1     {\ensuremath{A_{\mathrm{T}}^{#1}}\xspace}           

\def\C#1      {\ensuremath{\mathcal{C}_{#1}}\xspace}                       
\def\Cp#1     {\ensuremath{\mathcal{C}_{#1}^{'}}\xspace}                    
\def\Ceff#1   {\ensuremath{\mathcal{C}_{#1}^{\mathrm{(eff)}}}\xspace}        
\def\Cpeff#1  {\ensuremath{\mathcal{C}_{#1}^{'\mathrm{(eff)}}}\xspace}       
\def\Ope#1    {\ensuremath{\mathcal{O}_{#1}}\xspace}                       
\def\Opep#1   {\ensuremath{\mathcal{O}_{#1}^{'}}\xspace}                    

\newcommand{\nospaceunit}[1]{\ensuremath{\text{#1}}}       
\newcommand{\aunit}[1]{\ensuremath{\text{\,#1}}}       

\newcommand{\tev}{\aunit{Te\kern -0.1em V}\xspace}
\newcommand{\gev}{\aunit{Ge\kern -0.1em V}\xspace}
\newcommand{\mev}{\aunit{Me\kern -0.1em V}\xspace}
\newcommand{\kev}{\aunit{ke\kern -0.1em V}\xspace}
\newcommand{\ev}{\aunit{e\kern -0.1em V}\xspace}
\newcommand{\mevc}{\ensuremath{\aunit{Me\kern -0.1em V\!/}c}\xspace}
\newcommand{\gevc}{\ensuremath{\aunit{Ge\kern -0.1em V\!/}c}\xspace}
\newcommand{\mevcc}{\ensuremath{\aunit{Me\kern -0.1em V\!/}c^2}\xspace}
\newcommand{\gevcc}{\ensuremath{\aunit{Ge\kern -0.1em V\!/}c^2}\xspace}

\def\mum  {\ensuremath{\,\upmu\nospaceunit{m}}\xspace}

\def\fb   {\ensuremath{\aunit{fb}}\xspace}
\def\invfb   {\ensuremath{\fb^{-1}}\xspace}

\def\ps   {\ensuremath{\aunit{ps}}\xspace}
\def\fs   {\ensuremath{\aunit{fs}}\xspace}

\newcommand{\stat}{\aunit{(stat)}\xspace}
\newcommand{\syst}{\aunit{(syst)}\xspace}

\newcommand{\chisq}{\ensuremath{\chi^2}\xspace}

\newcommand{\chisqip}{\ensuremath{\chi^2_{\text{IP}}}\xspace}

\def\gsim{{~\raise.15em\hbox{$>$}\kern-.85em
          \lower.35em\hbox{$\sim$}~}\xspace}
\def\lsim{{~\raise.15em\hbox{$<$}\kern-.85em
          \lower.35em\hbox{$\sim$}~}\xspace}

\def\sPlot{\mbox{\em sPlot}\xspace}

\def\pt         {\ensuremath{p_{\mathrm{T}}}\xspace}

\def\ptot       {\ensuremath{p}\xspace}

\def\mrad{\aunit{mrad}}

\def\evtgen     {\mbox{\textsc{EvtGen}}\xspace}

\def\geant      {\mbox{\textsc{Geant4}}\xspace}

\def\photos     {\mbox{\textsc{Photos}}\xspace}

\def\pythia     {\mbox{\textsc{Pythia}}\xspace}

\def\tell1  {TELL1\xspace}
\def\ukl1   {UKL1\xspace}



%% file: fixed-symbols-def.tex
\usepackage{xspace} 
\usepackage{upgreek}

\def\DeltaOrDeltabar  {\kern 0.18em\optbar{\kern -0.18em \PDelta}{}\xspace}
\def\XiOrXibar        {\kern 0.18em\optbar{\kern -0.18em \PXi}{}\xspace}
\def\SigmaOrSigmabar  {\kern 0.18em\optbar{\kern -0.18em \PSigma}{}\xspace}
\def\OmegaOrOmegabar  {\kern 0.18em\optbar{\kern -0.18em \POmega}{}\xspace}

\def\Lcp          {{\ensuremath{\Lz^+_\cquark}}\xspace}

\def\Xicp         {{\ensuremath{\Xires^+_\cquark}}\xspace}

\def\Xiccbare     {{\ensuremath{\Xires_{\cquark\cquark}}}\xspace}
\def\Xiccpp       {{\ensuremath{\Xires^{++}_{\cquark\cquark}}}\xspace}

\def\Xiccp        {{\ensuremath{\Xires^{+}_{\cquark\cquark}}}\xspace}

\newcommand{\TeVnosp}{\ifthenelse{\boolean{inbibliography}}{\ensuremath{~T\kern -0.05em eV}}{\ensuremath{\mathrm{Te\kern -0.1em V}}}}
\newcommand{\GeVnosp}{\ensuremath{\mathrm{Ge\kern -0.1em V}}}
\newcommand{\MeVnosp}{\ensuremath{\mathrm{Me\kern -0.1em V}}}
\newcommand{\keVnosp}{\ensuremath{\mathrm{ke\kern -0.1em V}}}
\newcommand{\eVnosp}{\ensuremath{\mathrm{e\kern -0.1em V}}}
\newcommand{\GeVcnosp}{\ensuremath{{\mathrm{Ge\kern -0.1em V\!/}c}}}
\newcommand{\MeVcnosp}{\ensuremath{{\mathrm{Me\kern -0.1em V\!/}c}}}
\newcommand{\GeVccnosp}{\ensuremath{{\mathrm{Ge\kern -0.1em V\!/}c^2}}}
\newcommand{\MeVccnosp}{\ensuremath{{\mathrm{Me\kern -0.1em V\!/}c^2}}}
\newcommand{\GeVGeVccccnosp}{\ensuremath{{\mathrm{Ge\kern -0.1em V^2\!/}c^4}}}

\def\genxicctwo   {\mbox{\textsc{GenXicc2.0}}\xspace}

\def\Xiccbare     {{\ensuremath{\Xires_{\cquark\cquark}}}\xspace}


%% file: user-symbols-def.tex
\def\Xicc{{\ensuremath{\Xires_{\cquark\cquark}}}\xspace}
\def\Xiccp{{\ensuremath{\Xires_{\cquark\cquark}^{+}}}\xspace}
\def\Xiccpp{{\ensuremath{\Xires_{\cquark\cquark}^{++}}}\xspace}
\def\Omegaccbare{{\ensuremath{\Omegares_{\cquark\cquark}}}\xspace}
\def\Omegacc{{\ensuremath{\Omegares^{+}_{\cquark\cquark}}}\xspace}
\def\XiccppDecay{{\ensuremath{\Xiccpp\to\Lc\Km\pip\pip}}\xspace}

\def\LcDecay{{\ensuremath{\Lc\to p\Km\pip}}\xspace}

\def\logchisqip{{\ensuremath{\log_{10}(\chisqip)}}\xspace}

\newcommand{\xx}{\ensuremath{\kern 0.5em }}

\def\sPlot{\mbox{\em sPlot}\xspace}

\usepackage{multirow} 
\usepackage{booktabs} 
\usepackage{rotating}

%% file: title-LHCb-PAPER.tex

\begin{titlepage}
\pagenumbering{roman}

\vspace*{-1.5cm}
\centerline{\large EUROPEAN ORGANIZATION FOR NUCLEAR RESEARCH (CERN)}
\vspace*{1.5cm}
\noindent
\begin{tabular*}{\linewidth}{lc@{\extracolsep{\fill}}r@{\extracolsep{0pt}}}
\ifthenelse{\boolean{pdflatex}}
{\vspace*{-1.5cm}\mbox{\!\!\!\includegraphics[width=.14\textwidth]{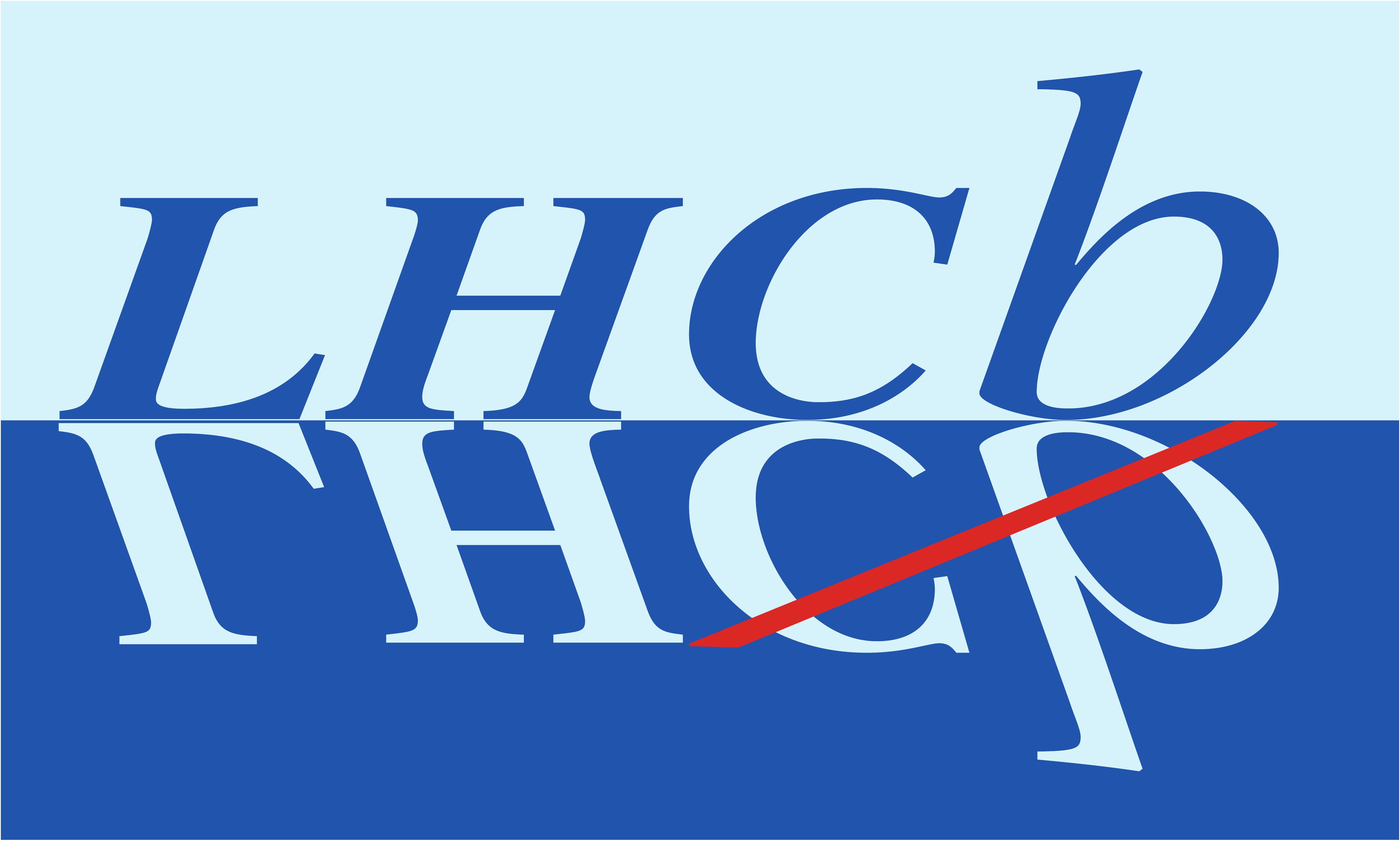}} & &}%
{\vspace*{-1.2cm}\mbox{\!\!\!\includegraphics[width=.12\textwidth]{figs/lhcb-logo.eps}} & &}%
\\
 & & CERN-EP-2019-199 \\  
 & & LHCb-PAPER-2019-029 \\  
 & & November 22, 2019 \\ 
 & & \\
\end{tabular*}

\vspace*{4.0cm}

{\normalfont\bfseries\boldmath\huge
\begin{center}
  \papertitle 
\end{center}
}

\vspace*{2.0cm}

\begin{center}
\paperauthors\footnote{Authors are listed at the end of this paper.}
\end{center}

\vspace{\fill}

\begin{abstract}
  \noindent
  A search for the doubly charmed baryon \Xiccp is performed through
  its decay to the $\Lc\Km\pip$ final state,
  using proton-proton collision data collected with the \lhcb detector
  at centre-of-mass energies of 7, 8 and 13\tev. 
  The data correspond to a total integrated luminosity of 9\invfb.
  No significant signal is observed in the mass range from 3.4 to 3.8\gevcc.
  Upper limits are set at $95\%$ credibility level 
  on the ratio of the \Xiccp production cross-section times 
  the branching fraction
  to that of
  \Lc and \Xiccpp baryons.
  The limits are determined as functions of the \Xiccp mass for different lifetime hypotheses, 
  in the rapidity range from 2.0 to 4.5
  and the transverse momentum range from 4 to 15\gevc.
\end{abstract}

\vspace*{2.0cm}

\begin{center}
  Published in Sci. China-Phys. Mech. Astron. 63, 221062 (2020) 
\end{center}

\vspace{\fill}

{\footnotesize 
\centerline{\copyright~\papercopyright. \href{\paperlicenceurl}{\paperlicence}.}}
\vspace*{2mm}

\end{titlepage}


\newpage
\setcounter{page}{2}
\mbox{~}
%
%
%
%

%% file: introduction.tex
\section{Introduction}%
\label{sec:introduction}

The constituent quark model~\cite{
  GellMann:1964nj,
  Zweig:352337,
  *Zweig:570209}
predicts the existence of 
weakly decaying doubly charmed baryons with
spin-parity $J^{P}=1/2^{+}$. These include
one isospin doublet \Xicc
($\Xiccp=\cquark\cquark\dquark$ and $\Xiccpp=\cquark\cquark\uquark$), 
and one isospin singlet \Omegaccbare ($\Omegacc=\cquark\cquark\squark$).
The masses of the two \Xicc states are predicted to be in the range from 
3500 to 3700\mevcc\cite{
  Kerbikov:1987vx,
  Fleck:1989mb,
  Chernyshev:1995gj,
  Gershtein:1998sx,*Gershtein:1998un,*Gershtein:2000nx,
  Itoh:2000um,
  Anikeev:2001rk,
  Kiselev:2001fw,
  Ebert:2002ig,
  Mathur:2002ce,
  He:2004px,
  Roberts:2007ni,
  Valcarce:2008dr,
  Zhang:2008rt,
  wang2010analysis,
  Aliev:2012ru,
  Namekawa:2013vu,
  Sun:2014aya,
  Karliner:2014gca,
  Brown:2014ena,
  Padmanath:2015jea,
  Bali:2015lka,
  Wei:2015gsa,
  Sun:2016wzh,
  Alexandrou:2017xwd,
  Liu:2017xzo,*Liu:2017frj},
with an isospin splitting of 
a few\:\mevcc\cite{Hwang:2008dj,Brodsky:2011zs,Karliner:2017gml}.
Predictions of the $\Xiccp$ lifetime span the range of 50 to 250\fs, 
while the \Xiccpp lifetime is predicted to be three to four times larger
due to the \mbox{$\W$-exchange} contribution in the \Xiccp decay
and the destructive Pauli interference in the \Xiccpp decay~\cite{
  Fleck:1989mb,
  guberina1999inclusive,
  Kiselev:1998sy,
  Likhoded:1999yv,
  Onishchenko:2000yp,
  Anikeev:2001rk,
  Kiselev:2001fw,
  hsi2008lifetime,
  Karliner:2014gca,
  Berezhnoy2016}.

Doubly charmed baryons have been searched for by
several experiments in the past decades. 
The SELEX collaboration reported the observation of the \Xiccp baryon decaying into \Lcp\Km\pip and $\proton \Dp \Km$ final states~\cite{
  Mattson:2002vu,
  Ocherashvili:2004hi},
using a 600\gevc charged hyperon beam impinging on a fixed target.
The mass of the \Xiccp baryon, averaged over the two decay modes, was found to be \mbox{$3518.7 \pm 1.7 \mevcc$}.
The lifetime was measured to be less than 33\fs at 90\% confidence level.
It was estimated that about 20\% of \Lc baryons in the SELEX experiment were produced from \Xiccp decays~\cite{Mattson:2002vu}.
Searches in different production environments
by the FOCUS~\cite{Ratti:2003ez},
\babar~\cite{Aubert:2006qw},
\lhcb~\cite{LHCb-PAPER-2013-049}
and \belle~\cite{Kato:2013ynr} experiments
did not confirm the SELEX results. 
Recently, the \Xiccpp baryon was observed by the LHCb experiment
in the $\Lcp \Km \pip \pip$ final state~\cite{LHCb-PAPER-2017-018},
and confirmed in the $\Xicp \pip$ final
state~\cite{LHCb-PAPER-2018-026}.
The weighted average of the \Xiccpp mass of the two decay modes was determined to be
$3621.24 \pm 0.65\stat \pm 0.31\syst \mevcc$~\cite{LHCb-PAPER-2018-026},
which is about $100\mevcc$ higher than the mass of the \Xiccp baryon reported by SELEX.
The lifetime of the \Xiccpp baryon was measured to be
$\ensuremath{0.256\,^{+0.024}_{-0.022}\stat\pm
0.014\syst\ps}$~\cite{LHCb-PAPER-2018-019},
which establishes its weakly decaying nature.
The $\Xiccpp\to\Dp\proton\Km\pip$ decay has been searched for by the LHCb
collaboration with a data sample corresponding to an integrated luminosity of 1.7\invfb, 
but no signal was found~\cite{LHCb-PAPER-2019-011}.

This paper presents the result of 
a search for the $\Xiccp$ baryon in the mass range from 3400 to 3800\mevcc,
where the $\Xiccp$ baryon is reconstructed through the $\Xiccp\to\Lc\Km\pip$, $\Lc \to \proton\Km\pip$ decay chain. 
The inclusion of charge-conjugate decay processes is implied throughout this paper. 
The data set comprises $pp$ collision data 
recorded with the \lhcb detector at centre-of-mass energies
$\sqrt{s}=7\tev$ in 2011, $\sqrt{s}=8\tev$ in 2012
and $\sqrt{s}=13\tev$ in 2015--2018,
corresponding to an integrated luminosity of 1.1$\invfb$, 2.1$\invfb$
and 5.9\invfb, respectively.
This data sample is about ten times larger than that of the
previous \Xiccp search by the \lhcb collaboration using only 2011 data~\cite{LHCb-PAPER-2013-049}.

The search was performed 
with the whole analysis procedure defined before 
inspecting the data in the 3400 to 3800\mevcc mass range.
The analysis strategy is defined as follows: first a search for a
$\Xiccp$ signal is performed and the significance of the signal as a
function of the $\Xiccp$ mass is evaluated; then if the global
significance, after considering the look-elsewhere effect,
is above 3 standard deviations, 
the $\Xiccp$ mass is measured;
otherwise, upper limits are set on the production rates
for different centre-of-mass energies.
Two sets of selections, with different multivariate classifiers and trigger requirements,  
denoted as \mbox{Selection A} and \mbox{Selection B}, are used in these two cases. 
\mbox{Selection A} is used in the signal search and is designed to maximise its sensitivity.
\mbox{Selection B} is optimised for setting upper limits on the ratio
of the $\Xiccp$ production rate to that of $\Xiccpp$ and $\Lcp$
baryons. It uses the same selection for $\Lcp$ baryons from $\Xicc$
decays and prompt $\Lcp$ baryons in order to have better control over
sources of systematic uncertainty on the ratio.
For the limit setting, only
the data recorded at $\sqrt{s}=8\tev$ in 2012
and at $\sqrt{s}=13\tev$ in 2016--2018 is used.
The 2015 data is excluded because there were significant variations
in trigger thresholds during this data-taking period, and because this
sample only accounts for 6\% of the $pp$ collision data at $\sqrt{s}=13\tev$.
The production ratio, $\mathcal{R}$, is defined as
\begin{equation}
  \label{eq:RLc}
  \mathcal{R}(\Lc) \equiv \frac{\sigma(\Xiccp)\times\BF(\Xiccp \to \Lc\Km\pip)}{\sigma(\Lc)}
\end{equation}
relative to the prompt $\Lcp$ baryons decaying to $\proton\Km\pip$, and 
\begin{equation}
  \label{eq:RXicc}
  \mathcal{R}(\Xiccpp) \equiv \frac{\sigma(\Xiccp)\times\BF(\Xiccp \to
  \Lc\Km\pip)}{\sigma(\Xiccpp)\times\BF(\Xiccpp \to \Lc\Km\pip\pip)}
\end{equation}
relative to the $\Xiccpp \to \Lc\Km\pip\pip$ decay,
where $\sigma$ is the production cross-section and
$\mathcal{B}$ is the decay branching fraction.
The determination of the ratio $\mathcal{R}(\Lc)$ allows a direct comparison with previous experiments,
while that of $\mathcal{R}(\Xiccpp)$ provides information 
about the ratio of the branching fractions of the 
$\Xiccp \to \Lc\Km\pip$ and $\Xiccpp \to \Lc\Km\pip\pip$ decays
assuming that the members of the isospin doublet have a similar production cross-section~\cite{Kiselev:2001fw, Ma:2003zk, Chang:2006eu}. 
The production ratios are evaluated as
\begin{equation}
  \mathcal{R} = \frac{\varepsilon_{\text{norm}}}{\varepsilon_{\text{sig}}}
                \frac{N_{\text{sig}}}{N_{\text{norm}}}
              \equiv \alpha N_{\text{sig}},
\end{equation}
where $\varepsilon_{\text{sig}}$ and $\varepsilon_{\text{norm}}$ 
refer to the selection efficiencies of the \Xiccp signal decay mode
and the \Lc or \Xiccpp normalisation decay modes respectively,
$N_{\text{sig}}$ and $N_{\text{norm}}$ are the corresponding yields, and $\alpha$
is the single-event sensitivity. 
Because the \Xiccp selection efficiency depends strongly on the lifetime,
limits on $\mathcal{R}(\Lc)$ and $\mathcal{R}(\Xiccpp)$
are quoted as functions of the \Xiccp signal mass 
for a discrete set of lifetime hypotheses.

%% file: detector.tex
\section{Detector and simulation}
\label{sec:Detector}

The \lhcb detector~\cite{LHCb-DP-2008-001,LHCb-DP-2014-002} is 
a single-arm forward
spectrometer covering the \mbox{pseudorapidity} range $2<\eta <5$,
designed for the study of particles containing \bquark or \cquark
quarks. The detector includes a high-precision tracking system
consisting of a silicon-strip vertex detector surrounding the $pp$
interaction region~\cite{LHCb-DP-2014-001},
a large-area silicon-strip detector located
upstream of a dipole magnet with a bending power of about
$4{\mathrm{\,Tm}}$, and three stations of silicon-strip detectors and straw
drift tubes~\cite{LHCb-DP-2013-003,LHCb-DP-2017-001}
placed downstream of the magnet.
The tracking system provides a measurement of the momentum, \ptot, of charged particles with
a relative uncertainty that varies from 0.5\% at low momentum to 1.0\% at 200\gevc.
The minimum distance of a track to a primary vertex (PV), the impact parameter (IP), 
is measured with a resolution of $(15+29/\pt)\mum$,
where \pt is the component of the momentum transverse to the beam, in\,\gevc.
Different types of charged hadrons are distinguished using information
from two ring-imaging Cherenkov detectors\cite{LHCb-DP-2012-003}. 
The online event selection is performed by a trigger~\cite{LHCb-DP-2012-004}, 
which consists of a hardware stage, based on information from the calorimeter and muon
systems, followed by a software stage, which applies a full event
reconstruction.

Simulated samples are required to develop the event selection and
to estimate the efficiency of the detector acceptance and the
imposed selection requirements.
Simulated $pp$ collisions are generated using
\pythia~\cite{Sjostrand:2007gs,*Sjostrand:2006za} 
with a specific \lhcb configuration~\cite{LHCb-PROC-2010-056}.  
A dedicated generator, \genxicctwo~\cite{Chang:2009va},
is used to simulate the \Xiccbare baryon production.
Decays of unstable particles
are described by \evtgen~\cite{Lange:2001uf}, in which final-state
radiation is generated using \photos~\cite{Golonka:2005pn}. The
interaction of the generated particles with the detector, and its response,
are implemented using the \geant
toolkit~\cite{Allison:2006ve, *Agostinelli:2002hh} as described in
Ref.~\cite{LHCb-PROC-2011-006}.
Unless otherwise stated,
simulated events are generated with a \Xicc mass of 3621\mevcc
and a $\Xiccp\,(\Xiccpp)$ lifetime of 80\fs (256\fs).

%% file: selection.tex
\section{Reconstruction and selection}%
\label{sec:reconstruction_and_selection}

For the \Xiccp signal and each of the normalisation modes,
\Lc candidates are reconstructed in the $\proton\Km\pip$ final state.
At least one of the three \Lc decay products is 
required to pass an inclusive software trigger,
which requires that a track with associated large transverse momentum is inconsistent with originating from any PV.
For data recorded at $\sqrt{s}=13\tev$,
at least one of the three \Lc decay products is 
required to pass a multivariate selection
applied at the software trigger level~\cite{pmlr-v14-gulin11a, LHCb-PROC-2015-018}.
The \chisqip is defined as the difference in \chisq
of the PV fit with and without the particle in question.
The PV of any single particle is defined to be that with respect to which
the particle has the smallest \chisqip.
Candidate $\Lc$ baryons are formed from the combination of three tracks of good quality that do not originate from any PV and have large transverse momentum.  
Particle identification (PID) requirements are imposed on all three
tracks to suppress combinatorial background and misidentified charm-meson decays.
The \Lc candidates are also required to
have a mass in the range from 2211 to 2362\mevcc.

The \Xiccp candidates are reconstructed 
by combining a \Lc candidate with two tracks,
identified as \Km and \pip mesons using PID information.
The kaon and pion tracks are required to 
have a large transverse momentum and a good track quality.
To suppress duplicate tracks,
the angle between each pair of the five final-state tracks with the same charge is
required to be larger than 0.5\mrad.
The \Xiccp candidate is required to have $\pt>4\gevc$ and to originate from a PV.
Similar requirements are imposed to reconstruct the \Xiccpp candidates
in the \Xiccpp normalisation mode,
with an additional charged pion in the final state.

Multivariate classifiers based on the
gradient boosted decision tree (BDTG)~\cite{Breiman,Hocker:2007ht,*TMVA4} are developed to further improve the signal purity. 
To train the classifier, 
simulated \Xiccp events are used as the signal sample and 
wrong-sign (WS) $\Lc\Km\pim$ combinations selected from the data sample
are used as the background sample.
For \mbox{Selection A},
the classifier is trained using candidates
with a $\Lc$ mass in the 
window of 2270 to 2306\mevcc
(corresponding to $\pm3$ times the resolution on the \Lc mass)
and a \Xiccp mass in the
signal search region.
Eighteen input variables that show good discrimination
for \Xiccp and intermediate \Lc candidates between signal and background samples
are used in the training.
These variables can be 
subdivided into two sets;
in the choice of 
the first set of variables, no strong assumptions are made
about the source of the \Lc candidates, while for the second set of variables the properties of the \Xiccp candidates
as the source of the \Lc candidates are exploited.
The first set of variables are:
the \chisq per degree of freedom of the \Lc vertex fit;
the \pt of the \Lc candidate and of its decay products;
and the flight-distance \chisq between the PV and the decay vertex
of the \Lc candidate. 
The second set of variables are:
the \chisq per degree of freedom of the \Xiccp vertex fit and of
the kinematic refit~\cite{Hulsbergen:2005pu} of the decay chain
requiring \Xiccp to originate from its PV;
the largest distance of closest approach (DOCA) between the decay products
of the \Xiccp candidate;
the \pt of the \Xiccp candidate, and of the kaon and pion from the \Xiccp decay;
the $\chisqip$ of the \Xiccp and \Lc candidates, and of the \Km and \pip mesons from the \Xiccp decay;
the angle between the momentum and displacement vector of the \Xiccp candidate;
and the flight-distance \chisq between the PV and the decay vertex
of the \Xiccp candidate.
For \mbox{Selection B},
the multivariate selection comprises two stages.
In the first stage,
one classifier is trained with $\Lcp$ signal in the simulated $\Xiccp$ sample and 
background candidates
in the \Lc mass sideband, 
and is applied to both the signal mode and the \Lc normalisation mode.
The same input variables are used as for the first set of variables in \mbox{Selection A},
with four additional variables that enhance the discriminating power:
the largest DOCA between the decay products
of the \Lc candidate
and the $\chisqip$ of the decay products of the \Lc candidate.
In the second stage,
another classifier is trained for the signal mode using candidates
in the mass window of the intermediate \Lc and the \Xiccp signal search region.
Candidates used in the training are also required to
pass a BDTG response threshold of the first classifier.
The input variables are those from the second set of \mbox{Selection A}
with an additional variable, 
the angle between the momentum and displacement vector of the \Lc candidate.

The thresholds of the BDTG responses for both \mbox{Selection A} and B
are determined by maximising the expected value of the figure of merit
$\varepsilon/\left(\frac{5}{2} + \sqrt{N_B}\right)$~\cite{Punzi:2003bu},
where $\varepsilon$ is the estimated signal efficiency, 
$5/2$ corresponds to 5 standard deviations in a Gaussian significance test, 
and $N_B$ the expected number of background candidates
under the signal peak.
The quantity $N_B$ is estimated with the WS control sample
in the mass region of $\pm12.5\mevcc$ around the known \Xiccpp mass~\cite{PDG2018},
taking into account the difference of the background level for the signal sample
and the WS control sample.
The performance of the BDTG classifier is tested and found to be stable against the $\Xiccp$ lifetimes in the range from 40 to 120\fs. 
Following the same procedure,
a two-stage multivariate selection is developed for the \Xiccpp normalisation mode.

Events that pass the multivariate selection may contain 
more than one \Xiccp candidate in the search region although the probability to produce more than one \Xiccp is small.
According to studies of simulated decays and the WS control sample,
multiple candidates in the same event do not form a peaking background except for one
case in which the candidates are obtained from the same five final-state
tracks, but with two tracks interchanged (e.g. the \Km from the \Lc decay and 
the \Km from the \Xiccp decay).
In this case,
only one candidate is chosen randomly.

For \mbox{Selection B},
an additional hardware trigger requirement is imposed on candidates of both the signal
and the normalisation mode 
to minimise systematic differences in efficiency between the modes. This hardware trigger requirement 
selects candidates in which at least one of the three \Lc decay products 
deposits high transverse energy in the calorimeters. Finally, 
\Xiccp baryon candidates in the signal mode and \Lc and \Xiccpp
baryons in the normalisation modes
are required to be reconstructed in the fiducial region of
rapidity $2.0<y<4.5$ and transverse momentum $4<\pt<15\gevc$.

%% file: yield.tex
\section{Yield measurements}
\label{sec:yield}
\begin{figure}[tb]
  \centering
  \includegraphics[width=0.49\linewidth]{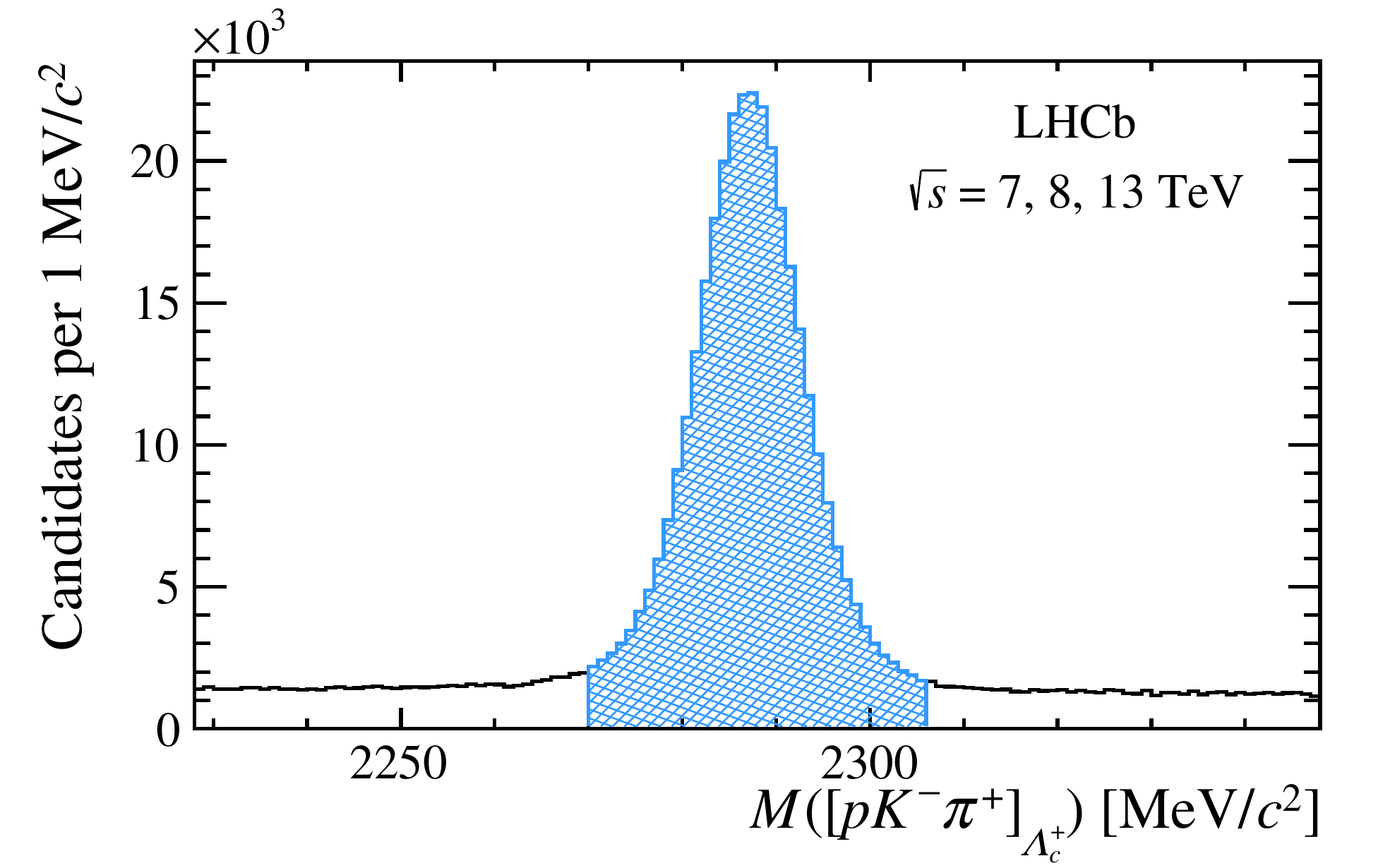}
  \includegraphics[width=0.49\linewidth]{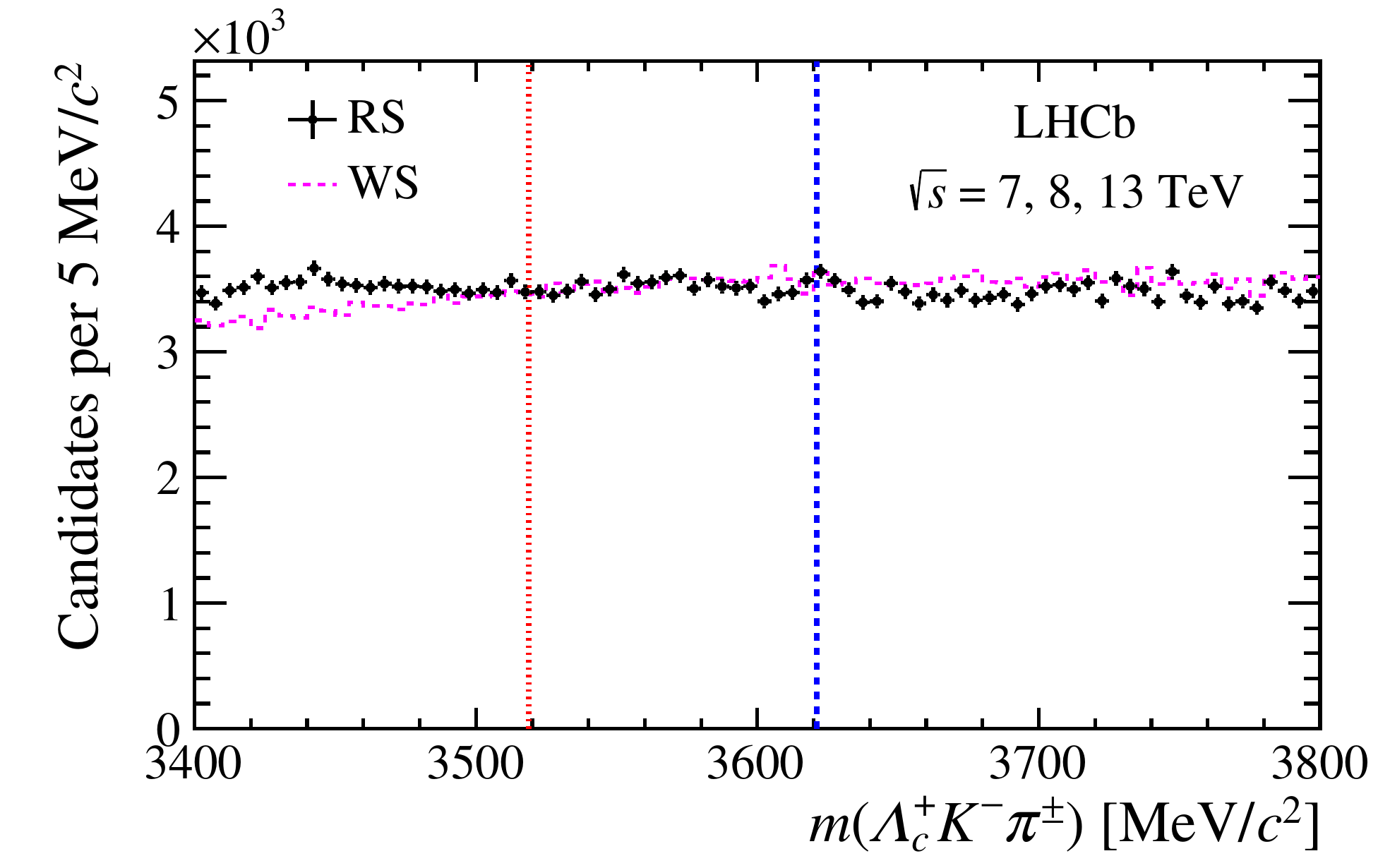}
  \caption{Mass distributions of the (left) intermediate \Lc and (right) \Xiccp candidates
  for the full data sample.
  Selection A is applied, including the \Lc mass requirement,
  indicated by the cross-hatched region in the left plot,
  of $2270\mevcc<M([\proton\Km\pip]_{\Lc})<2306\mevcc$.
  The right-sign (RS) $m(\Lc\Km\pip)$ distribution is shown in the right plot,
  along with the wrong-sign (WS) $m(\Lc\Km\pim)$ distribution
  normalised to have the same area.
  The dotted red line at 3518.7\mevcc indicates the mass of the \Xiccp baryon reported by SELEX~\cite{Ocherashvili:2004hi}
  and the dashed blue line at 3621.2\mevcc indicates the mass of the isospin partner, the \Xiccpp baryon~\cite{LHCb-PAPER-2018-026}.}
  \label{fig:unblind_Xiccp}
\end{figure}

\begin{figure}[tb]
  \centering
  \includegraphics[width=0.6\linewidth]{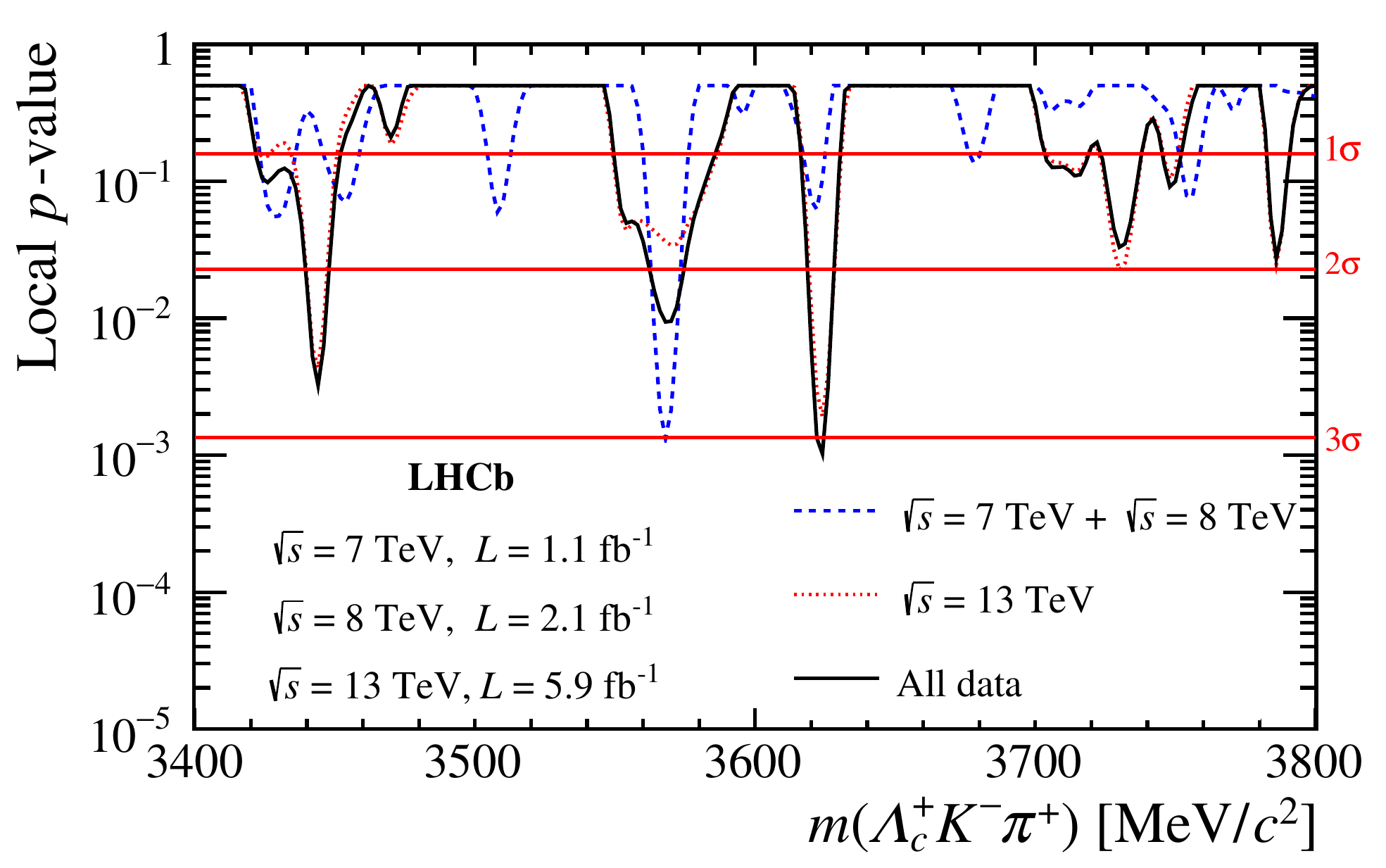}
  \caption{Local $p$-value (statistical only) at different $\Xiccp$ mass values
  evaluated with the likelihood-ratio test, for the data sets recorded at $\sqrt{s}=7\tev$, $\sqrt{s}=8\tev$ and $\sqrt{s}=13\tev$.
  Selection A is applied, including the \Lc mass requirement
  of $2270\mevcc<M([\proton\Km\pip]_{\Lc})<2306\mevcc$.}
  \label{fig:pvalue}
\end{figure}

Selection A described above is applied to the full data sample.
Figure~\ref{fig:unblind_Xiccp} shows the $M([\proton\Km\pip]_{\Lc})$ and
$m(\Lc\Km\pip)$ distributions in the \Lc mass range from
2270\mevcc to 2306\mevcc. The quantity $m(\Lc\Km\pip)$ is defined as
\begin{equation}
  \label{eq:mXicc}
  m(\Lc\Km\pip) \equiv M([\proton\Km\pip]_{\Lc}\Km\pip) 
                       - M([\proton\Km\pip]_{\Lc}) 
                       + M_{\mathrm{PDG}}(\Lc),
\end{equation}
where $M([\proton\Km\pip]_{\Lc}\Km\pip)$ is 
the reconstructed mass of the \Xiccp candidate,
$M([\proton\Km\pip]_{\Lc})$ is the reconstructed mass of the \Lc candidate,
and $M_{\mathrm{PDG}}(\Lc)$ is the known value of the \Lc mass~\cite{PDG2018}.
As a comparison, the $m(\Lc\Km\pim)$ distribution of the WS control sample
is also shown in the right plot of Fig.~\ref{fig:unblind_Xiccp}.
The dotted red line indicates the mass of the \Xiccp baryon reported by SELEX~\cite{
  Mattson:2002vu,
  Ocherashvili:2004hi},
and the dashed blue line refers to the mass of the \Xiccpp baryon~\cite{
  LHCb-PAPER-2017-018,
  LHCb-PAPER-2018-026}.
The small enhancement below 3500\mevcc, compared to the WS sample, is due to  partially reconstructed $\Xiccpp$ decays. 
There is no excess near a mass of 3520\mevcc.
A small enhancement is seen near a mass of 3620\mevcc.
To determine the statistical significance of this enhancement,
an extended unbinned maximum-likelihood fit is performed to
the $m(\Lc\Km\pip)$ distribution.
The signal component is described with the sum of
a Gaussian function and
a modified Gaussian function with power-law tails on both sides~\cite{Skwarnicki:1986xj}.
The parameters of the signal model are fixed from simulation
except for the common peak position of the two functions that is allowed to vary freely in the fit.
The background component is described by
a second-order Chebyshev polynomial with all parameters free.
A local $p$-value is evaluated with the likelihood ratio test for rejection of the background-only hypothesis assuming a positive signal~\cite{Wilks:1938dza,Narsky:2000}
and is shown in Fig.~\ref{fig:pvalue}.
The largest local significance, corresponding to $3.1$
standard deviations ($2.7$ standard deviations after considering systematic uncertainties), occurs around 3620\mevcc.
Taking into account the look-elsewhere effect
in the mass range of 3500\mevcc to 3700\mevcc
following Ref.~\cite{Gross:2010qma},
the global $p$-value is $4.2\times10^{-2}$,
corresponding to a significance of $1.7$ standard deviations.
Since no excess above 3 standard deviations is observed,
upper limits on the production ratios are set using the 
 data recorded at $\sqrt{s}=8\tev$ in 2012
and at $\sqrt{s}=13\tev$ in 2016--2018
after applying \mbox{Selection B}.

\begin{figure}[tb]
  \centering
  \includegraphics[width=0.45\linewidth]{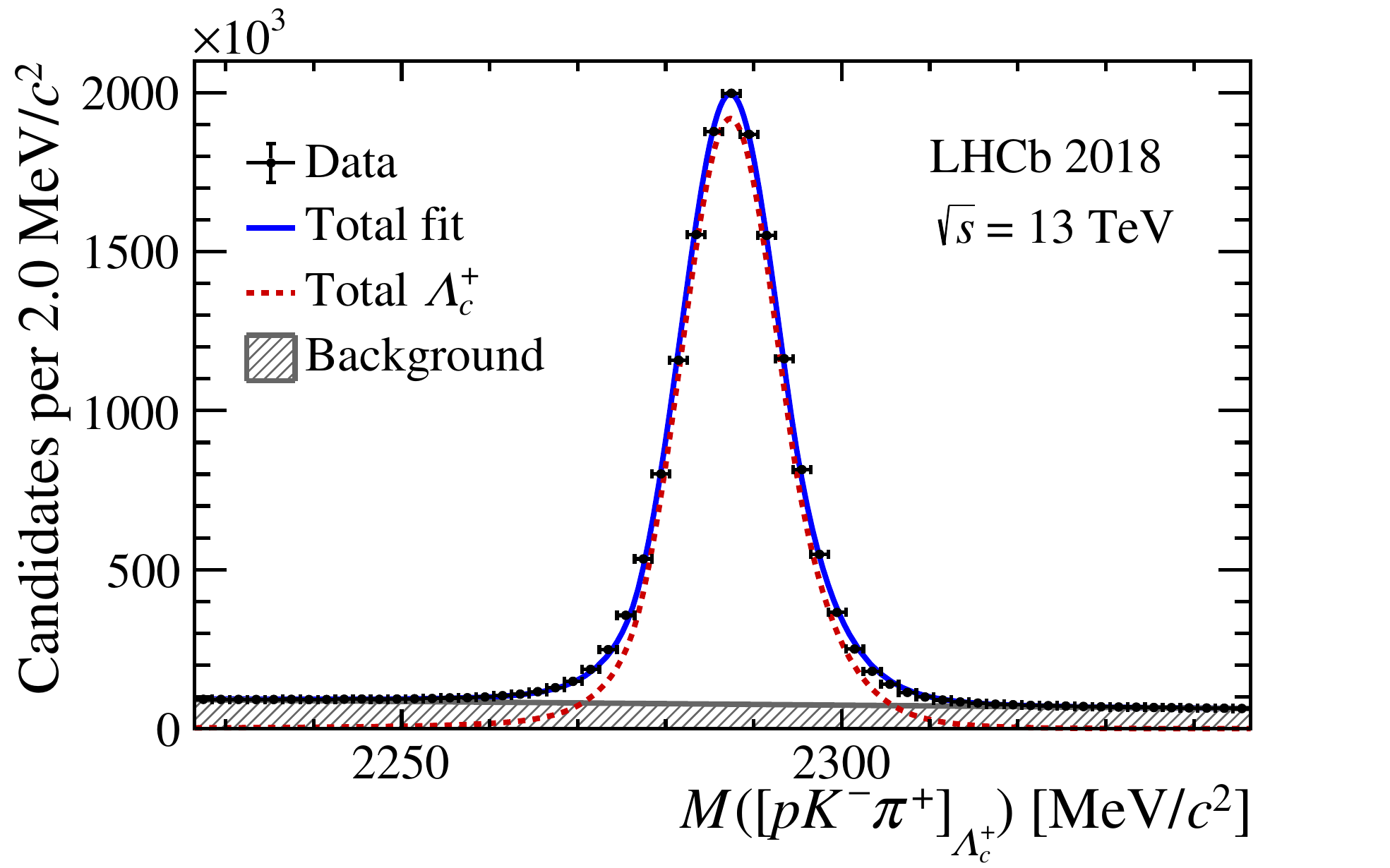}
  \includegraphics[width=0.45\linewidth]{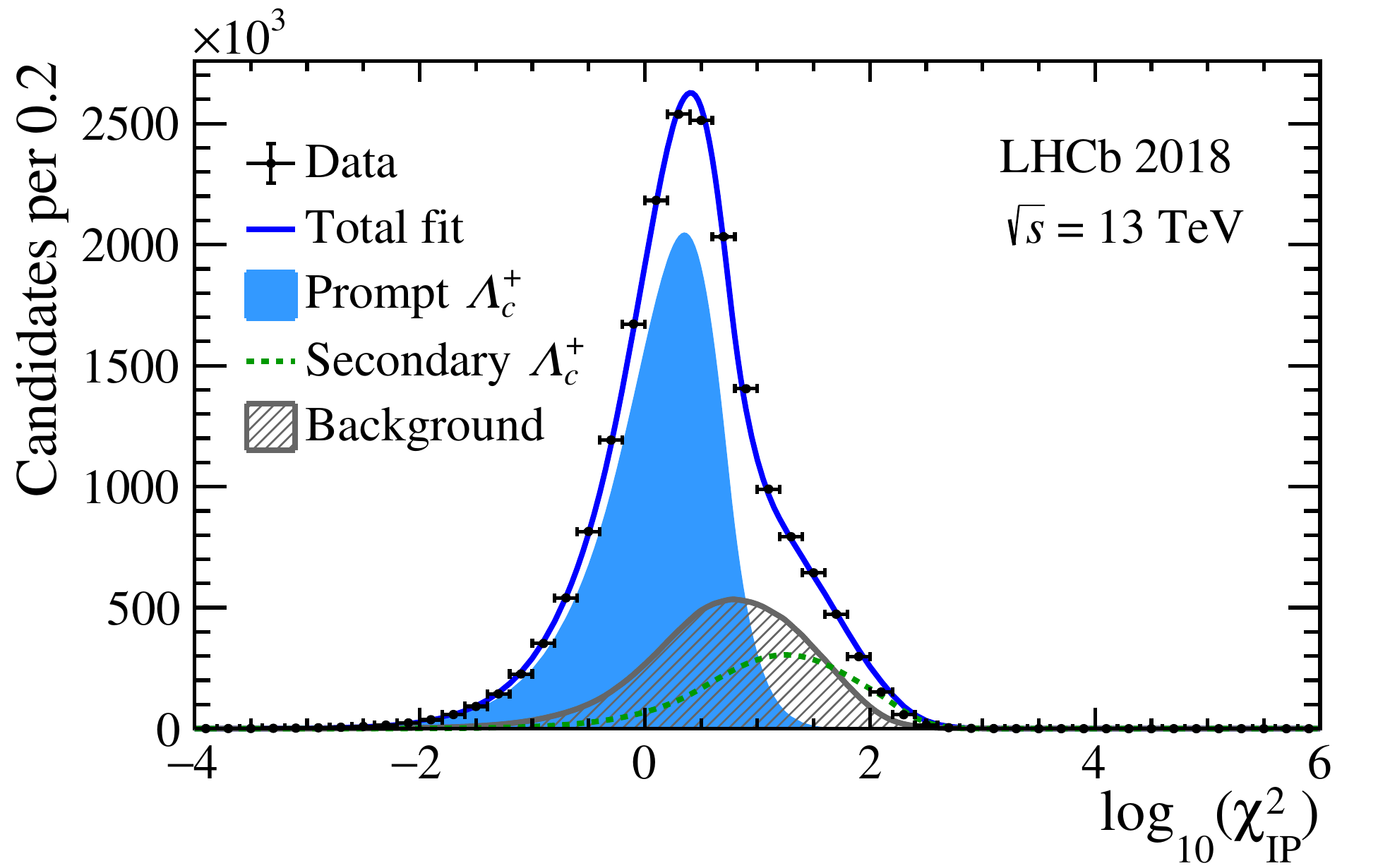}\\
  \caption{Distributions of (left) $M([\proton\Km\pip]_{\Lc})$ and (right) \logchisqip of the selected $\Lc$ candidates
  with associated fit results for the 2018 data set.}
  \label{fig:yield_Lc}
\end{figure}

To measure the production ratios,
it is necessary to determine the yields of the normalisation modes.
The yield determination procedure of the prompt \Lc decays
is complicated by the substantial secondary \Lc contribution
from $\bquark$-hadron decays,
and is done in two steps.
First, the total number of \Lc candidates is 
determined with an extended unbinned maximum-likelihood fit to
the $M([\proton\Km\pip]_{\Lc})$ distribution.
Then, a fit to
the \logchisqip distribution is performed to
discriminate between prompt and secondary \Lc candidates.
Information from the \Lc mass fit is used to constrain
the total number of \Lc candidates.
The shapes of the prompt and secondary \logchisqip distributions
are described by
a Bukin function~\cite{bukin2007fitting}.
The shape parameters of the prompt and secondary components
are determined from simulation,
except for the mean and the width parameters of the Bukin function, which are allowed to vary in the fit. 
The background component is described by
a nonparametric function generated
using the data from the \Lc mass sideband regions.
As an illustration,
the $M([\proton\Km\pip]_{\Lc})$
and \logchisqip distributions
of the $\Lc$ normalisation mode
candidates in the 2018 data set are shown in Fig.~\ref{fig:yield_Lc}.
The prompt \Lc yields are summarised in Table~\ref{tab:yield_summary}.

\begin{figure}[tb]
  \centering
  \includegraphics[width=0.6\linewidth]{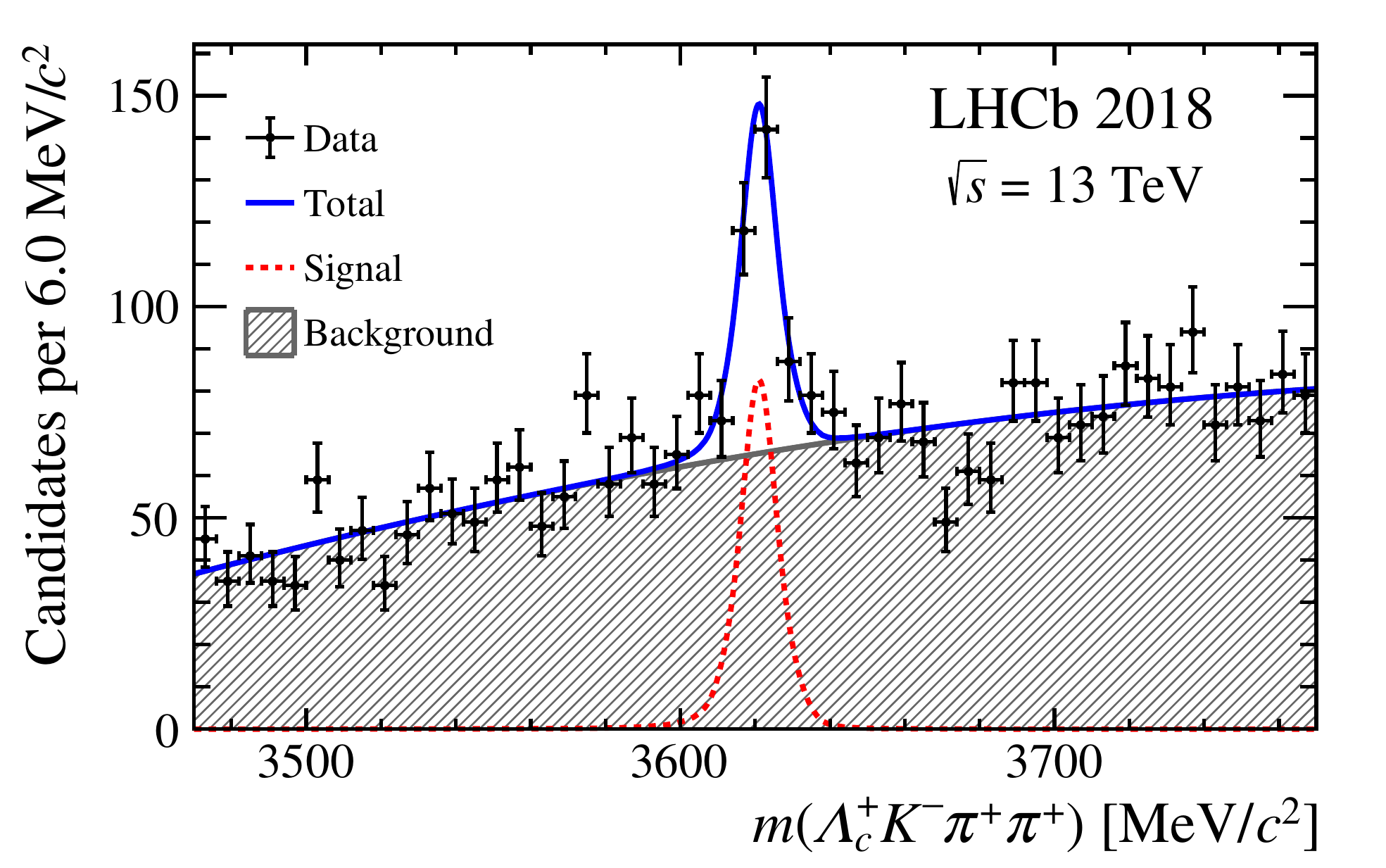}
  \caption{Mass distribution of $\Xiccpp$ candidates in the 
  2018 data set. The result of a fit to the distribution is shown. }
  \label{fig:yield_xicc}
\end{figure}

To determine the \Xiccpp yield, 
an extended unbinned maximum-likelihood fit is
performed to the $m(\Lc\Km\pip\pip)$ distribution,
which is defined in a similar way to Eq.~\ref{eq:mXicc}.
The same signal and background parameterisations are used as for the signal mode.
For the data sample recorded at $\sqrt{s}=13\tev$,
a simultaneous fit is performed to the \mbox{$m(\Lc\Km\pip\pip)$} distributions
of the candidates in the 
2016, 2017 and 2018 data sets with the shared mean and resolution parameter.
As an illustration,
the $m(\Lc\Km\pip\pip)$ distribution for the 2018 data set
is shown in Fig.~\ref{fig:yield_xicc} along with the associated fit result.
The \Xiccpp yields are summarised in Table~\ref{tab:yield_summary}.
\begin{table}[bt]
  \centering
  \caption{Signal yields for prompt $\Lc\to\proton\Km\pip$ and $\Xiccpp\to\Lc\Km\pip\pip$ normalisation modes,
   split by data-taking period. The integrated luminosity $\mathcal{L}$ is also shown for each data-taking period.}
  \label{tab:yield_summary}
  \begin{tabular}{ccrr}
    \toprule
    Period   & $\mathcal{L}$ [\invfb]& $N(\Lc)$ [$\times 10^{3}$] & $N(\Xiccpp)$ \\
    \midrule
    2012&   2.1&  $1175.3\pm2.5$   & $38 \pm10$ \\
    2016&   1.7&  $7339\pm\;12$   & $121\pm19$ \\
    2017&   1.7&  $9883\pm \phantom{0.}9$   & $153\pm22$ \\
    2018&   2.2&  $11184\pm\; 13$  & $188\pm24$ \\
    \bottomrule
  \end{tabular}
\end{table}

%% file: efficiency.tex
\section{Efficiency ratio measurement}%
\label{sec:efficiency_ratio}

 To set upper limits on the production ratios, the efficiency ratio $\eps_{\rm norm}/\eps_{\rm sig}$ is determined from simulation.
The signal efficiency is estimated with mass and lifetime
hypotheses of $m(\Xiccp)=3621\mevcc$ and $\tau(\Xiccp)=80\fs$.
The kinematic distribution of the \Xiccp baryon is assumed to be the same
as for its  isospin partner \Xiccpp and the \pt distribution of simulated
\Xiccp decays is corrected according to the data-simulation discrepancy
observed in the \Xiccpp normalisation mode.
The Dalitz distributions of the simulated \Lc decays are
corrected to match the distribution observed in  background-subtracted data, obtained using 
 the \sPlot technique~\cite{Pivk:2004ty}.
Corrections are applied to the tracking efficiency and PID response of the simulated samples using calibration data samples~\cite{LHCb-DP-2013-002,LHCb-PUB-2016-021,LHCb-DP-2018-001}. 
The efficiency ratio obtained for the \Lc and \Xiccpp normalisation modes
and for different data-taking years are summarised in Table~\ref{tab:eff_ratio},
where the uncertainties are due to the limited sizes of the simulated
samples. The increase in the efficiency ratio of the \Xiccpp normalisation mode in 2017--2018 compared to that in 2016
is due to the improvement of the online event selection following the observation of the \Xiccpp baryon.
\begin{table}[tb]
  \centering
  \caption{Efficiency ratios between the normalisation and signal modes for 
  different data-taking periods.
  The uncertainties are due to the limited size of the simulated samples.}
  \label{tab:eff_ratio}
    \begin{tabular}{l c c c c}
    \toprule
    & 2012 & 2016 & 2017 & 2018 \\
    \midrule
    $\varepsilon_{\mathrm{norm}}(\Lc)/\varepsilon_{\mathrm{sig}}$     &  54  $\pm$  17    & 22.0  $\pm$  1.9\;\;   & 22.4  $\pm$  1.3\;\;    & 26.1  $\pm$  1.8\;\;  \\
    $\varepsilon_{\mathrm{norm}}(\Xiccpp)/\varepsilon_{\mathrm{sig}}$ & 2.1  $\pm$  0.7   & 1.17  $\pm$  0.11  & 1.91  $\pm$  0.11  & 1.99  $\pm$  0.12 \\
    \bottomrule
  \end{tabular}
\end{table}

The signal efficiency of the event selection has a strong dependence on
the \Xiccp lifetime. To estimate the efficiency for
other lifetime hypotheses, the decay time of the simulated \Xiccp events
are weighted to have different exponential distributions
and the efficiency is re-calculated.
A discrete set of hypotheses (40\fs, 80\fs, 120\fs, and 160\fs)
is motivated by the measured \Xiccpp lifetime of 256\fs~\cite{LHCb-PAPER-2018-019} and 
the expectation that the \Xiccp lifetime is 
three to four times smaller than that of the \Xiccpp baryon~\cite{
  Fleck:1989mb,
  guberina1999inclusive,
  Kiselev:1998sy,
  Likhoded:1999yv,
  Onishchenko:2000yp,
  Anikeev:2001rk,
  Kiselev:2001fw,
  hsi2008lifetime,
  Karliner:2014gca,
  Berezhnoy2016}.
Combining the yields of the normalisation modes obtained in the previous section,
the values of the single-event sensitivity of the 
\Lc and \Xiccpp modes 
for several lifetime hypotheses are shown in 
Table~\ref{tab:alpha_Lc} and Table~\ref{tab:alpha_Xicc},  respectively.
The uncertainties on the single-event sensitivities are due to the limited sizes of the simulated samples
and the statistical uncertainties on the measured yields.
\begin{table}[tb]
  \centering
  \caption{Single-event sensitivity of the \Lc normalisation mode
  $\alpha(\Lc)$ [$\times 10^{-5}$]
  for different lifetime hypotheses of the $\Xiccp$ baryon in the different data-taking years.
  The uncertainties are due to the limited sizes of the simulated samples
  and the statistical uncertainties on the measured \Lc baryon yields.}
\label{tab:alpha_Lc}
\begin{tabular}{c cccc} 
\toprule 
 & $\tau=40\fs$ & $\tau=80\fs$ & $\tau=120\fs$ & $\tau=160\fs$ \\
\midrule 
2012& 14.2 $\pm$ 4.8\;\; & 4.6 $\pm$ 1.4& 2.65 $\pm$ 0.77& 1.91 $\pm$ 0.53\\ 
2016& 0.60 $\pm$ 0.08& 0.29 $\pm$ 0.02& 0.20 $\pm$ 0.01& 0.16 $\pm$ 0.01\\ 
2017& 0.46 $\pm$ 0.04& 0.23 $\pm$ 0.01& 0.15 $\pm$ 0.01& 0.12 $\pm$ 0.01\\ 
2018& 0.52 $\pm$ 0.04& 0.23 $\pm$ 0.02& 0.15 $\pm$ 0.01& 0.11 $\pm$ 0.01\\ 
\bottomrule 
\end{tabular} 
\end{table}

\begin{table}[tb]
  \centering
  \caption{Single-event sensitivity of the \Xiccpp normalisation mode 
  $\alpha(\Xiccpp)$ [$\times 10^{-2}$]
  for different lifetime hypotheses of the $\Xiccp$ baryon in the different data-taking years.
  The uncertainties are due to the limited size of the simulated samples
  and the statistical uncertainty on the measured \Xiccpp baryon yield.}
\label{tab:alpha_Xicc}
\begin{tabular}{c cccc} 
\toprule 
 & $\tau=40\fs$ & $\tau=80\fs$ & $\tau=120\fs$ & $\tau=160\fs$ \\ 
\midrule 
2012& 16.7 $\pm$ 7.1\phantom{7}& 5.4 $\pm$ 2.2& 3.1 $\pm$ 1.2& 2.3 $\pm$ 0.8\\ 
2016& 1.96 $\pm$ 0.42& 0.96 $\pm$ 0.18& 0.65 $\pm$ 0.12& 0.52 $\pm$ 0.09\\ 
2017& 2.51 $\pm$ 0.42& 1.25 $\pm$ 0.19& 0.84 $\pm$ 0.13& 0.69 $\pm$ 0.11\\ 
2018& 2.36 $\pm$ 0.34& 1.06 $\pm$ 0.15& 0.68 $\pm$ 0.10& 0.52 $\pm$ 0.08\\ 
\bottomrule 
\end{tabular} 
\end{table}

The efficiency could depend on the \Xiccp mass,
since it affects the kinematic distributions of the decay products of the \Xiccp baryon.
To test other mass hypotheses,
two simulated samples are generated with
$m(\Xiccp)=3518.7\mevcc$ and $m(\Xiccp)=3700.0\mevcc$.
The \pt distributions of the three decay products of the \Xiccp
in the simulated sample with $m(\Xiccp)=3621.4\mevcc$
are weighted to match those in the other mass hypotheses,
and the efficiency is re-calculated with the weighted sample.
Despite the variations of individual efficiency components,
the total efficiency is found to be independent of such variations.
The mass dependence can be effectively ignored for the evaluation of 
the single-event sensitivities.

%% file: systematics.tex
\section{Systematic uncertainties}

The systematic uncertainties on the measured production ratio $\mathcal{R}$ are
presented in Table~\ref{tab:ratio_sys_summary}. The total systematic
uncertainty is calculated as the quadratic sum of the individual
uncertainties, assuming all sources to be independent.

The largest systematic uncertainty arises from the evaluation of the
efficiency of the hardware-trigger requirement. The cancellation of
the hardware-trigger efficiencies in the ratio of the signal and the
normalisation decay channels is studied with the \Lc and \Lb control
samples, using a tag-and-probe method~\cite{LHCb-DP-2012-004}. 
The difference between the efficiency ratio in data
and in simulation is assigned as systematic uncertainty. 

The systematic uncertainty on the yield determination is evaluated by varying the choice of the model used to fit the data. 
For the \XiccppDecay decay,
an alternative model is used where 
the signal is described by the sum of two Gaussian functions and 
the background is described by a second-order polynomial function.
For the \mbox{\LcDecay} normalisation mode, the yield of the prompt $\Lc$ is
determined by the fit to the \logchisqip distribution. The
uncertainty on the determined signal yield may arise from signal
modelling and the limited size of the sample in the background region
of the \Lc invariant mass used to model the background.
For the signal modelling,
a bifurcated Gaussian with an exponential tail is used. The effect of the background is evaluated through the use of pseudoexperiments. The background population in 
each bin of the
\logchisqip
template is fluctuated randomly, and the fit procedure is repeated.

The PID efficiency is determined in bins of particle momentum
and pseudorapidity using calibration data samples.
The effect of the limited size of the calibration samples is studied
with a large number of pseudoexperiments and that of the binning
scheme is studied by increasing the number of bins
by a factor of two.
The sum in quadrature of these effects is taken as systematic uncertainty arising from PID efficiency.

The tracking efficiency is corrected with calibration data samples~\cite{LHCb-DP-2013-002}.
There are three sources of systematic uncertainties related to this correction.
The first is due to the limited size of the calibration samples and
is estimated with pseudoexperiments.
The second is due to the calibration method and an uncertainty of 0.8\% (0.4\%) per
track is assigned for the 13\tev (7\tev) data~\cite{LHCb-DP-2013-002}.
The third is due to the imperfect knowledge of the material budget in the detector.
The above contributions to the systematic uncertainty are summed in quadrature.

\begin{table}
  \centering
  \caption{Summary of the systematic uncertainties of the production ratio measurement.}
  \label{tab:ratio_sys_summary}
  \begin{tabular}{l rrrr}
    \toprule
    \multirow{2}*{Source} & \multicolumn{2}{c}{$\sqrt{s}=8\tev$}&  \multicolumn{2}{c}{$\sqrt{s}=13\tev$}\\
    \cline{2-3}
    \cline{4-5}
    ~ & $\mathcal{R}(\Lc)$ & $\mathcal{R}(\Xiccpp)$ & $\mathcal{R}(\Lc)$ & $\mathcal{R}(\Xiccpp)$ \\
    \midrule
    Trigger efficiency   &11.7\% & 17.7\%  & 4.9\% & 11.2\%  \\
    Yield measurement    &5.8\%  & 8.9\%   & 0.6\% & 0.4\%  \\
    PID efficiency       &2.5\%  & 4.6\%   & 0.9\% & 0.8\%  \\
    Tracking             &4.3\%  & 2.6\%   & 4.4\% & 3.1\%  \\
    \midrule
    Total& 14.0\% & 20.5\% & 6.7\% &  11.7\%\\
    \bottomrule
  \end{tabular}
\end{table}

%% file: upper_limit.tex
\section{Upper limits}%
\label{sec:results}

\begin{figure}[tb]
  \centering
  \includegraphics[width=0.49\linewidth]{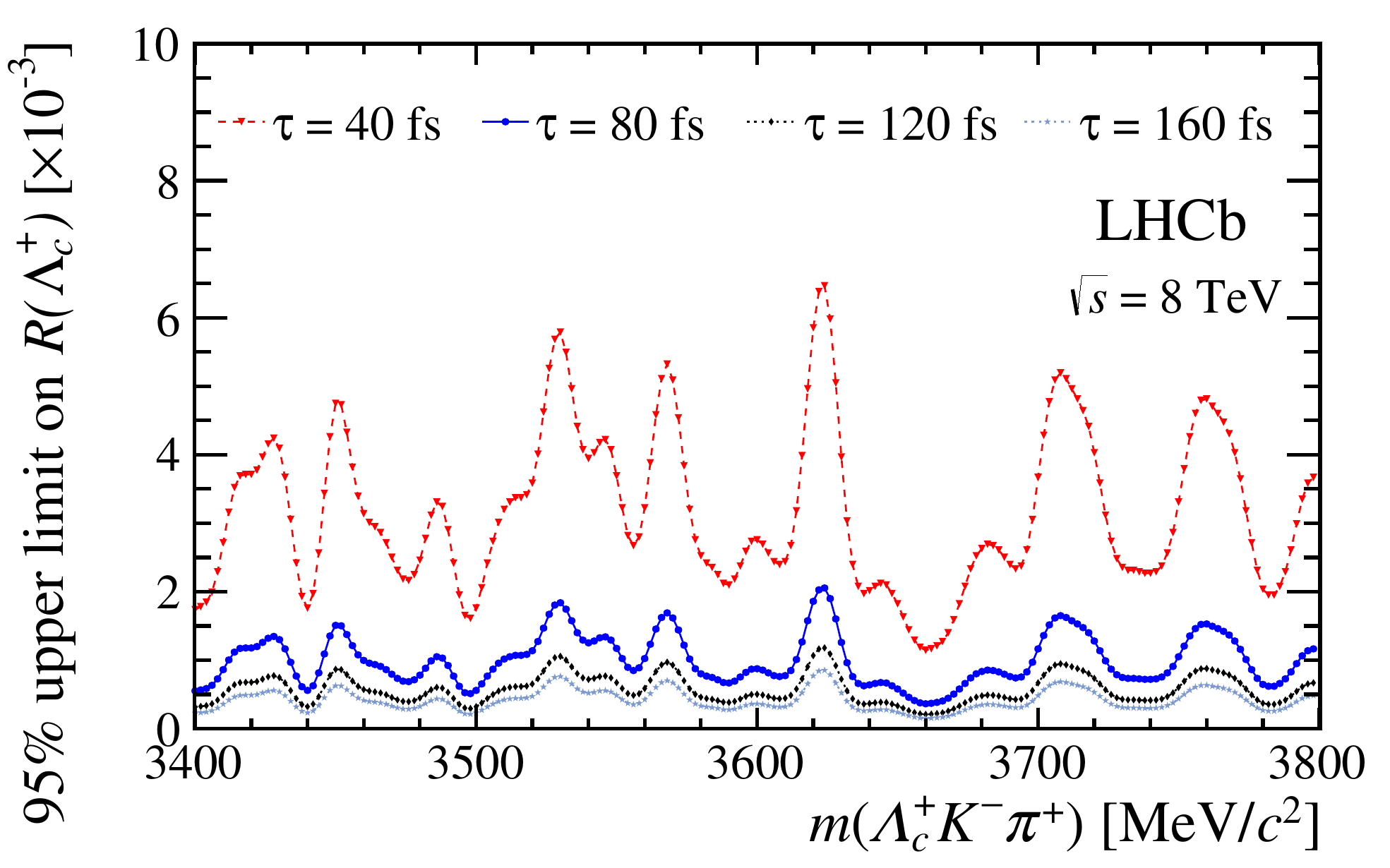}
  \includegraphics[width=0.49\linewidth]{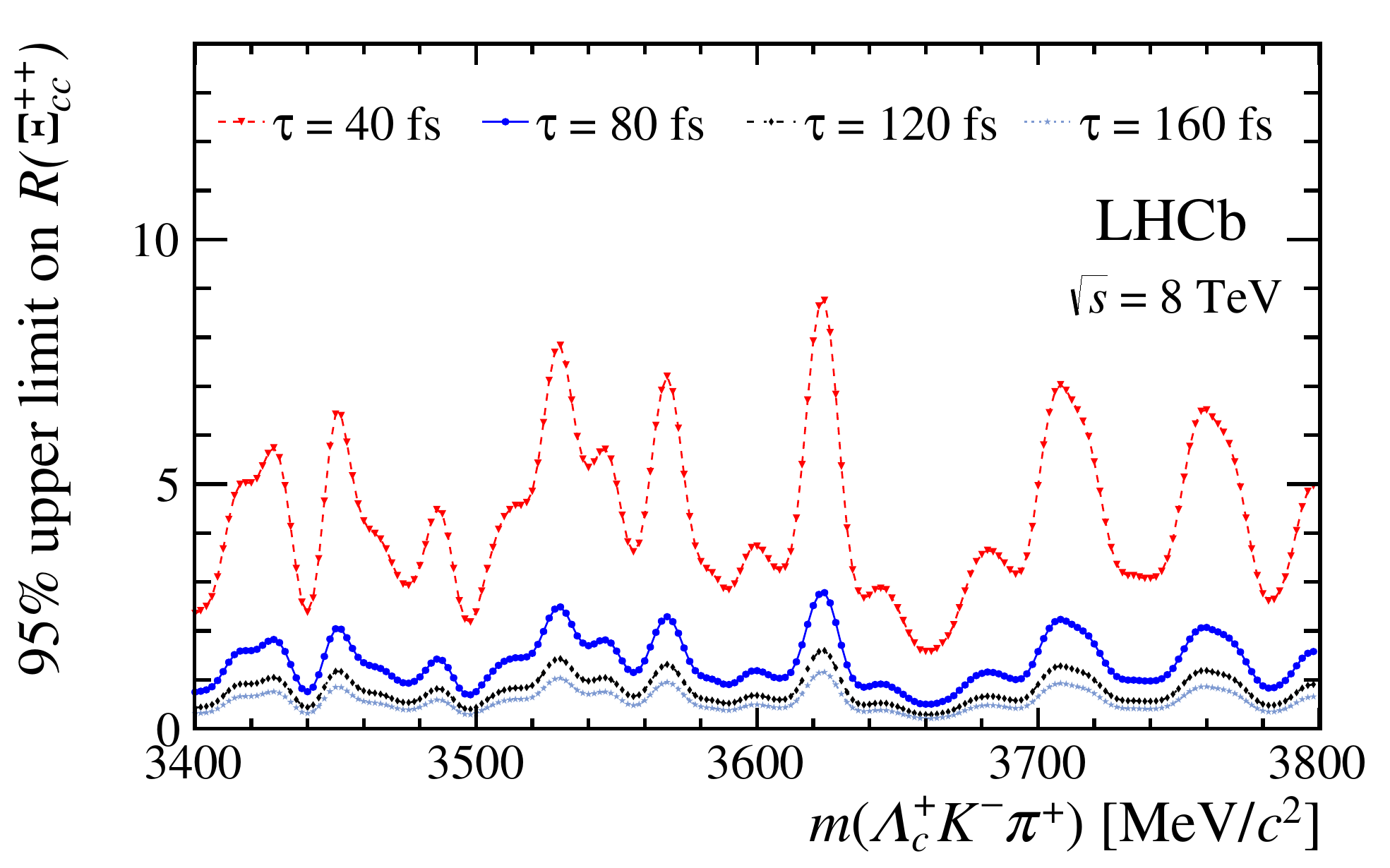}
  \caption{Upper limits on (left) $\mathcal{R}(\Lc)$
  and (right) $\mathcal{R}(\Xiccpp)$
  at 95\% credibility level as a function of $m(\Lc\Km\pip)$
  at $\sqrt{s}=8\tev$ for four $\Xiccp$ lifetime hypotheses.}
  \label{fig:mass_scan_tau_run1}
\end{figure}

\begin{figure}[tb]
  \centering
  \includegraphics[width=0.49\linewidth]{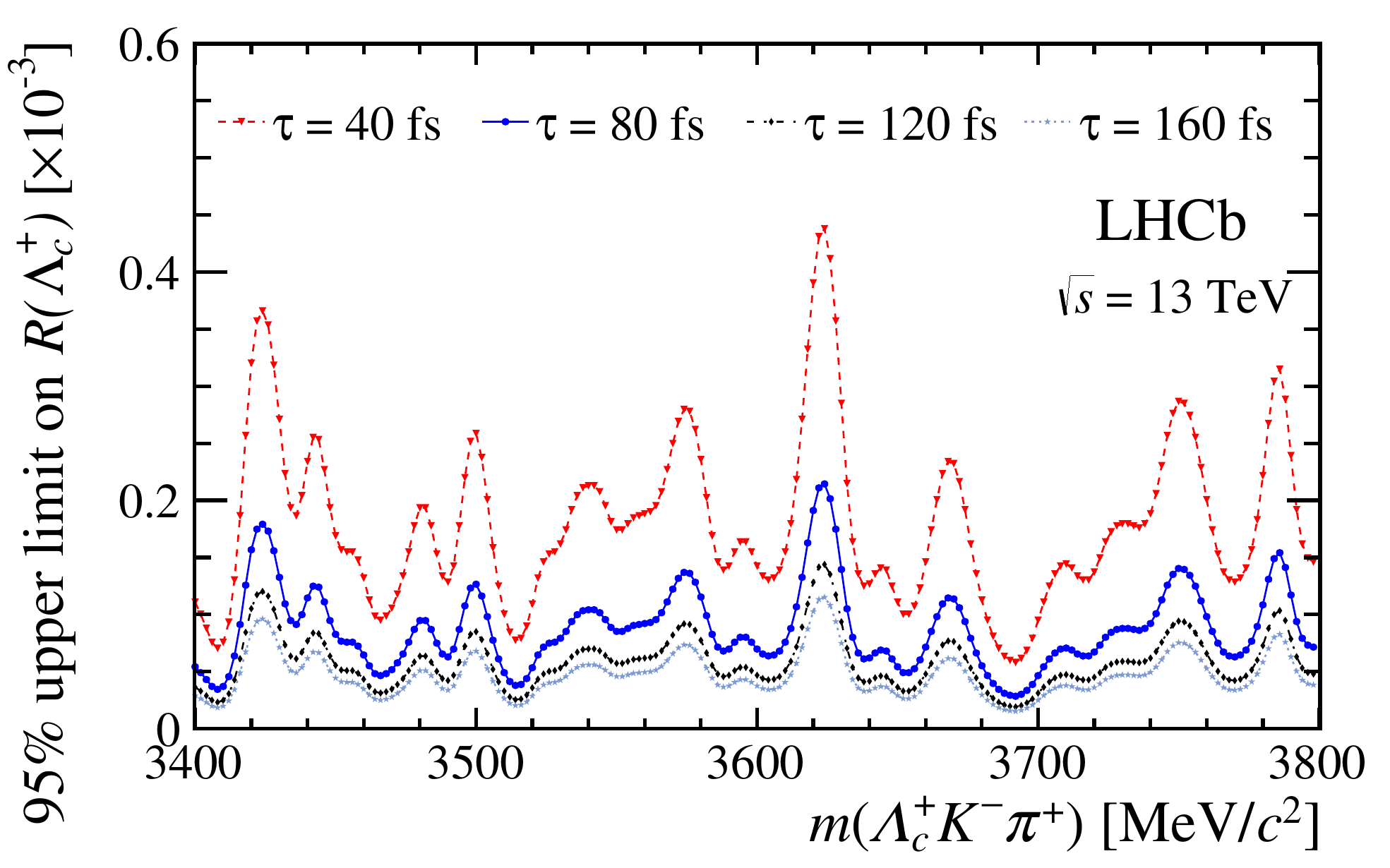}
  \includegraphics[width=0.49\linewidth]{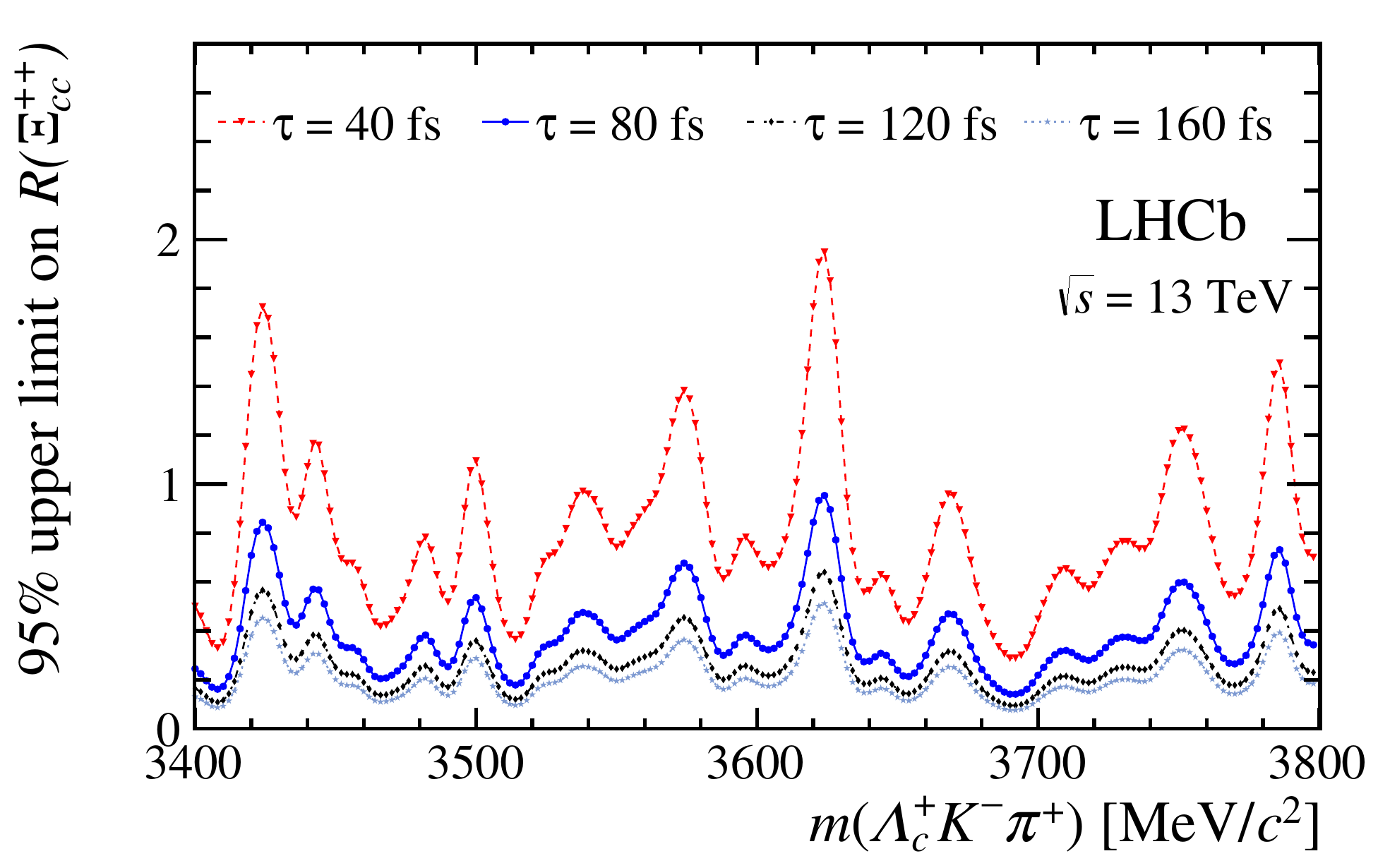}
  \caption{Upper limits on (left) $\mathcal{R}(\Lc)$
  (right) $\mathcal{R}(\Xiccpp)$
  at 95\% credibility level as a function of $m(\Lc\Km\pip)$ at
  $\sqrt{s}=13\tev$, 
  for four $\Xiccp$ lifetime hypotheses.}
  \label{fig:mass_scan_tau_run2}
\end{figure}

Upper limits at 95\% credibility level are set
on the production ratio $\mathcal{R}(\Lc)$ and $\mathcal{R}(\Xiccpp)$
at centre-of-mass energies $\sqrt{s}=8\tev$
and $\sqrt{s}=13\tev$, in the fiducial region of
rapidity $2.0<y<4.5$ and transverse momentum $4<\pt<15\gevc$.
Upper limits are calculated 
in 2.5\mevcc intervals over the $m(\Lc\Km\pip)$ mass range of 3400 to 3800\mevcc
for the four different lifetime hypotheses.
For each fixed value of the \Xiccp mass and lifetime,
the likelihood profile $\mathcal{L}(\mathcal{R})$ 
is determined as a function of $\mathcal{R}$.
The likelihood profile for the data recorded at $\sqrt{s}=13\tev$
is obtained with a simultaneous fit to the $m(\Lc\Km\pip)$ distributions
 using the same fit model as described in Sec.~\ref{sec:yield}.
Then the likelihood profile
$\mathcal{L}(\mathcal{R})$ is convolved with a Gaussian distribution whose width is equal to the square root of the quadratic combination of the 
statistical and systematic uncertainty on the single-event sensitivity.
The upper limit at 95\% credibility level is defined as
the value of $\mathcal{R}$ at which the integral starting from zero
equals 95\% of the total area under the curve.
Figures~\ref{fig:mass_scan_tau_run1} and~\ref{fig:mass_scan_tau_run2} 
show the 95\% credibility level upper limits
at centre-of-mass energies of $\sqrt{s}=8\tev$
and $\sqrt{s}=13\tev$, respectively.

%% file: conclusions.tex
\section{Conclusion}%
\label{sec:conclusion}

A search for the doubly charmed baryon \Xiccp is performed through
its decay to $\Lc\Km\pip$,
with the $pp$ collision data collected by the \lhcb experiment
at centre-of-mass energies of 7, 8 and 13\tev,
corresponding to an integrated luminosity of 9\invfb.
No significant signal is observed in the mass range from 3.4 to 3.8\gevcc.
Upper limits are set at $95\%$ credibility level
on the ratio of the \Xiccp production cross-section times
the branching fraction
to that of the
\Lc and \Xiccpp baryons.
The limits are determined as functions of the \Xiccp mass for different lifetime hypotheses, 
in the rapidity range from 2.0 to 4.5
and the transverse momentum range from 4 to 15\gevc.
The upper limit on the production ratio $R(\Lc)$ $(R(\Xiccpp))$
depends strongly on the considered mass and lifetime of the \Xiccp baryon, varying from
$0.45 \times 10^{-3}$ ($2.0$) for 40\fs to 
$0.12 \times 10^{-3}$ ($0.5$) for 160\fs, as summarised in Table~\ref{tab:ul_summary}.
The upper limits on $R(\Lc)$ are improved by one order of magnitude compared to the previous LHCb search~\cite{LHCb-PAPER-2013-049} 
and are significantly below the value reported by SELEX~\cite{Mattson:2002vu}, albeit in a different production environment.
Future searches by the \lhcb experiment with further improved trigger conditions, additional
$\Xiccp$ decay modes, and larger data samples should significantly increase the \Xiccp signal sensitivity. 

\begin{table}[tb]
  \centering
  \caption{Summary of the largest upper limits on production ratios at 95\% credibility level
  for four lifetime hypotheses and different centre-of-mass energies.}
  \label{tab:ul_summary}
  \begin{tabular}{rcccc}
    \toprule
    \multirow{2}*{Lifetime} & \multicolumn{2}{c}{$\sqrt{s}=8\tev$}&  \multicolumn{2}{c}{$\sqrt{s}=13\tev$}\\
    \cline{2-3}
    \cline{4-5}
    ~ & $\mathcal{R}(\Lc)$ [$\times 10^{-3}$] & $\mathcal{R}(\Xiccpp)$ & $\mathcal{R}(\Lc)$ [$\times 10^{-3}$] & $\mathcal{R}(\Xiccpp)$ \\
    \midrule
    40\fs   &6.5 & 8.8  & 0.45 & 2.0  \\
    80\fs   &2.1 & 2.8  & 0.22 & 1.0  \\
    120\fs  &1.2 & 1.6  & 0.15 & 0.6  \\
    160\fs  &0.9 & 1.2  & 0.12 & 0.5  \\  
    \bottomrule
  \end{tabular}
\end{table}

%% file: acknowledgements.tex
\section*{Acknowledgements}
\noindent 
We thank Chao-Hsi Chang, Cai-Dian L\"u, Wei Wang, Xing-Gang Wu, and
Fu-Sheng Yu for frequent and interesting discussions on the production
and decays of double-heavy-flavor baryons.
We express our gratitude to our colleagues in the CERN
accelerator departments for the excellent performance of the LHC. We
thank the technical and administrative staff at the LHCb
institutes. We acknowledge support from CERN and from the national
agencies: CAPES, CNPq, FAPERJ and FINEP (Brazil); MOST and NSFC
(China); CNRS/IN2P3 (France); BMBF, DFG and MPG (Germany); INFN
(Italy); NWO (Netherlands); MNiSW and NCN (Poland); MEN/IFA
(Romania); MinES and FASO (Russia); MinECo (Spain); SNSF and SER
(Switzerland); NASU (Ukraine); STFC (United Kingdom); NSF (USA).  We
acknowledge the computing resources that are provided by CERN, IN2P3
(France), KIT and DESY (Germany), INFN (Italy), SURF (Netherlands),
PIC (Spain), GridPP (United Kingdom), RRCKI and Yandex
LLC (Russia), CSCS (Switzerland), IFIN-HH (Romania), CBPF (Brazil),
PL-GRID (Poland) and OSC (USA). We are indebted to the communities
behind the multiple open-source software packages on which we depend.
Individual groups or members have received support from AvH Foundation
(Germany), EPLANET, Marie Sk\l{}odowska-Curie Actions and ERC
(European Union), ANR, Labex P2IO and OCEVU, and R\'{e}gion
Auvergne-Rh\^{o}ne-Alpes (France), Key Research Program of Frontier
Sciences of CAS, CAS PIFI, and the Thousand Talents Program (China),
RFBR, RSF and Yandex LLC (Russia), GVA, XuntaGal and GENCAT (Spain),
Herchel Smith Fund, the Royal Society, the English-Speaking Union and
the Leverhulme Trust (United Kingdom).

%% file: LHCb_Authorship_30-Jul-2019.tex
\centerline
{\large\bf LHCb collaboration}
\begin
{flushleft}
\small
R.~Aaij$^{31}$,
C.~Abell{\'a}n~Beteta$^{49}$,
T.~Ackernley$^{59}$,
B.~Adeva$^{45}$,
M.~Adinolfi$^{53}$,
H.~Afsharnia$^{9}$,
C.A.~Aidala$^{79}$,
S.~Aiola$^{25}$,
Z.~Ajaltouni$^{9}$,
S.~Akar$^{64}$
% ...

{\footnotesize \it

% ...
$ ^{9}$Universit{\'e} Clermont Auvergne, CNRS/IN2P3, LPC, Clermont-Ferrand, France\\
% ...
$ ^{25}$INFN Sezione di Milano, Milano, Italy\\
% ...
$ ^{31}$Nikhef National Institute for Subatomic Physics, Amsterdam, Netherlands\\
% ...
$ ^{45}$Instituto Galego de F{\'\i}sica de Altas Enerx{\'\i}as (IGFAE), Universidade de Santiago de Compostela, Santiago de Compostela, Spain\\
% ...
$ ^{49}$Physik-Institut, Universit{\"a}t Z{\"u}rich, Z{\"u}rich, Switzerland\\
% ...
$ ^{53}$H.H. Wills Physics Laboratory, University of Bristol, Bristol, United Kingdom\\
% ...
$ ^{59}$Oliver Lodge Laboratory, University of Liverpool, Liverpool, United Kingdom\\
% ...
$ ^{64}$University of Cincinnati, Cincinnati, OH, United States\\
% ...
$ ^{67}$Syracuse University, Syracuse, NY, United States\\
% ...
$ ^{79}$University of Michigan, Ann Arbor, United States, associated to $^{67}$\\
\bigskip
% ...
}
\end{flushleft}